\documentclass[a4paper,oneside,12pt]{elsarticle}
\usepackage{amsthm}
\usepackage{amsmath}
\usepackage{graphicx}%
\usepackage{amsfonts}%
\usepackage{amssymb}
 \usepackage{verbatim}
 \usepackage{float}
\usepackage{rotating}
\usepackage{amsmath,amsthm,amssymb}

\newtheorem{defi}{Definition}

\newtheorem{prop}{Proposition}
\newtheorem{rem}{Remark}

\begin{document}
%\tableofcontents

\title{\huge{\textbf{Regime Switching Stochastic Volatility with Perturbation Based Option Pricing}}\\
\large{\textbf{by\\
Sovan Mitra}}} \maketitle
\section*{\large{Abstract}}
Volatility modelling has become a significant area of research
within Financial Mathematics. Wiener process driven stochastic
volatility models have become popular due their consistency with
theoretical arguments and empirical observations. However such
models lack the ability to take into account long term and
fundamental economic factors e.g. credit crunch.

Regime switching models with mean reverting  stochastic volatility
are a new class of stochastic volatility models that capture both
short and long term characteristics. We propose a new general
method of pricing options for these new class of stochastic
volatility models using Fouque's perturbation based option pricing
method.

Using empirical data, we compare our option pricing method to
Black-Scholes and Fouque's standard option pricing method and show
that our pricing method provides lower relative error compared to
the other two methods.
\\\\
\textbf{Key words}: Stochastic volatility, option pricing,
perturbation theory.

\line(1,0){400}
%\newpage
%%%%%%%%%%%%%%%%%%%%%%%%%%%%%%%%%%%%%%%%%%%%%%%%%%%%%%%%%%%%%%%%%%%%%%%%%%%%%%%%%%%%%%%%%%%%%%%%%%%%%%%%%%%%%%%%%%%%%%%%%%%%%%
\begin{comment}
\begin{center}
\begin{tabular}{c}
\\
\\
\\
\\
\\
\\
\\
\\
{\bf \large{``Risk comes from not knowing what you're doing",}}
\\
{\bf \large{Warren Buffett.}}
\\
\end{tabular}
\end{center}

\newpage
\end{comment}
%%%%%%%%%%%%%%%%%%%%%%%%%%%%%%%%%%%%%%%%%%%%%%%%%%%%%%%%%%%%%%%%%%%%%%%%%%%%%%%%%%%%%%%%%%%%%%%%%%%%%%%%%%%%%%%%%%%%%%%%%%%%%%

\section{Introduction and Outline}
Volatility modelling has become a significant area of research
within Financial Mathematics: it helps us understand price
dynamics since it is one of the key variables in a stochastic
differential equation. Volatility is the only variable in the
Black-Scholes option pricing equation that is unobservable, hence
crucial to option pricing.  Finally, volatility has a wide range
of industrial applications from pricing exotic derivatives to
asset pricing models \cite{rebonato2004vac}. Shiller \cite{UseVol}
also claims volatility can be used as a measure of market
efficiency.

With volatility modelling, stochastic volatility has achieved
growing importance as they capture a richer set of empirical and
theoretical characteristics compared to other volatility models
\cite{musiela2005mmf}:

\begin{itemize}
\item firstly, stochastic volatility models generate return
distributions similar to what is empirically observed.
%\cite{fouque2000dfm},42.
For example, the return distribution has a fatter left tail and
peakedness
 compared to normal distributions, with tail asymmetry controlled by
 $\rho$ \cite{durham2007smm}.

\item Secondly, Renault and Touzi \cite{renault1996oha} proved
 volatility that is stochastic and $\rho$=0 always produces implied volatilities that smile (note
that volatility smiles do not necessarily imply volatility is
stochastic).

\item Thirdly, historic volatility shows significantly higher
variability than would be expected from local or time dependent
volatility, which could be better explained by a stochastic
process. A particular case in point is the dramatic change in
volatility during the 1987 October crash (Schwert
\cite{schwert1990sva} gives an empirical study on this). Finally,
stochastic volatility accounts for the volatility's empirical
dependence on the time scale measured, which should not occur
under local or time dependent volatility.
%In particular if volatility is a stochastic mean reverting
%process, we would expect volatility to become more Gaussian as the
%time scale is increased (Gaussian aggregation);  this is
%empirically exhibited by stock prices (see section \ref{Empirical
%Characteristics of Volatility}) and will be proven in section
%\ref{Stochastic Volatility and the Driving Process}.
%A brief proof is given in \cite{fouque2000dfm,p52-3}.
%For $\rho \neq$0 no analytical conclusion can be drawn about the
%implied volatility relation. The reader is referred to
%\cite{fouque2000dfm} chapter 2 %p53-5
%for more information.
%It is interesting to note that modelling volatility as stochastic
%is not noticeable by visual inspection of the stock price process
%-SEE $http://www.xplore-stat.de/tutorials/stfhtmlnode46.html$

\end{itemize}

Wiener process driven stochastic volatility models capture price
and volatility dynamics more successfully compared to local and
time dependent volatility models. However, for longer term
dynamics and fundamental economic changes (e.g. ``credit crunch"),
no mechanism existed to address the change in volatility dynamics
and it has been empirically shown that volatility is related to
long term and fundamental conditions. To model volatility (and
stock prices) in continuous time but also capture long term and
fundamental factors requires regime switching models with mean
reverting  stochastic volatility. This is a currently growing area
of stochastic volatility models.

At present, there is no method for pricing options for volatility
governed by regime switching with mean reverting stochastic
volatility. Using Fouque's \cite{fouque2000dfm} perturbation
approach to stochastic volatility option pricing, we show how  it
is possible to obtain option prices. Additionally, the option
pricing method is possible for any mean reverting process. We
demonstrate our method with empirical results from S\&P 500 index
options.

The outline of the paper is as follows. Firstly we give a
literature review of volatility models and the option pricing
methods. We then explain Fouque's option pricing method under
standard mean reverting stochastic volatility and then show how it
can be applied to regime switching with meaning reverting
stochastic volatility. We finally provide numerical experiments,
demonstrating that our method of option pricing provides more
accurate prices compared to standard Fouque option pricing
(without regimes).

\section{Literature Review on Volatility Models}
\label{Volatility Modelling Literature Review chapter}  Bachelier
\cite{bachelier1900tst} proposed the first model for stock prices.
Bachelier reasoned that investing was theoretically a ``fair game"
in the sense that statistically one could neither profit nor lose
from it. Hence Bachelier included the Wiener process to
incorporate the random nature of stock prices. Osborne
\cite{osborne1959bms} conducted empirical work supporting
Bachelier's model. Samuelson \cite{samuelson1965ppa} continued the
constant volatility model under the Geometric Brownian Motion
stock price model on the basis of economic justifications.

Over time, empirical data and theoretical arguments found constant
volatility to be inconsistent with the observed market behaviour
(such as the leverage effect). A plot of the empirical daily
volatility of the S\&P 500 index clearly shows volatility is far
from constant. This consequently led to the development of dynamic
volatility modelling, which we will now discuss.

\subsection{Time Dependent Volatility Model}\label{Time Dependent
Volatility Model}
%\subsection{General Time Dependent Models}
%$http://www.xplore-stat.de/tutorials/stfhtmlnode52.html$\\
It was empirically observed that implied volatility varied with an
option's expiration date. Consequently, a straight forward
extension proposed to the constant volatility model was time
dependent
volatility modelling \cite{wilmott1998dta}:%p289
 \begin{eqnarray}
dX/X=\mu dt+\sigma(t)dW.
\end{eqnarray}
Merton \cite{merton1973tro} was the first to propose a formula for
pricing options under time dependent volatility.
The option price associated with X is still calculated by the standard Black-Scholes %wilmott
formula  except we set $\sigma={\sigma}_{c}$ where:
%NOTE: \sigma_{c}=\overline{\sigma^{2}} in Fouque p51
%\cite{fouque2000dfm,p39}:
\begin{eqnarray}\label{sigma_c}
\sigma_{c}=\sqrt{\dfrac{1}{T-t}\int^{T}_{t}\sigma^2(\tau)d\tau},
\end{eqnarray}
i.e. $d_{1}$ and $d_{2}$ in the Black-Scholes equation become:
\begin{eqnarray}
d_{1} &=& \dfrac{log \left(\dfrac{X}{K}\right) +\mu(T-t)+\dfrac{1}{2}\int^{T}_{t} \sigma^2(\tau) d\tau}{\sqrt{\int^{T}_{t}\sigma^2(\tau)d\tau}},\\
d_{2} &=& \dfrac{log \left(\dfrac{X}{K}\right)
+\mu(T-t)-\dfrac{1}{2}\int^{T}_{t} \sigma^2(\tau)
d\tau}{\sqrt{\int^{T}_{t}\sigma^2(\tau)d\tau}}.
\end{eqnarray}
The equation (\ref{sigma_c}) converts $\sigma(t)$ to its constant
volatility equivalent $\sigma_{c}$ over time period t to T. The
distribution of X(t) is given by:% \cite{fouque2000dfm,p39}
\begin{eqnarray}
log(X_{T}/X_{t})\sim \mathcal{N}\left((\mu-
\frac{1}{2}\sigma_{c}^{2})(T-t),\sigma_{c}^{2}(T-t)\right).
\end{eqnarray}
Note that the constant volatility $\sigma_{c}$ changes in value as
t and T change. This property enables time dependent volatility to
account for empirically observed implied volatilities increasing
with time (for a given strike).

\subsection{Local Volatility Model: Definition and Characteristics}
In explaining the empirical characteristics of volatility, a time
dependent volatility model was found to be insufficient. For
instance, time dependent volatility  did not explain the
volatility smile, nor the leverage effect, since volatility cannot
vary with price. Therefore volatility as a function of price (and
optionally time) was proposed, that is \textit{local volatility}
is
$\sigma=f(X,t)$: %(also known as restricted volatility and forward (forward) volatility \cite{wilmott1998dta,p292}):
\begin{eqnarray}
dX/X=\mu dt+\sigma(X,t)dW.
\end{eqnarray}
The term ``local" arises from knowing volatility with certainty
``locally" or when X is known -for a stochastic volatility model
we never know the volatility with certainty.

The advantages of local volatility models are that firstly, no
additional (or untradable) source of randomness is introduced into
the model.  Hence the models are complete, unlike stochastic
volatility models. It is theoretically possible to perfectly hedge
contingent claims. Secondly, local volatility models can be also
calibrated to perfectly fit empirically observed implied
volatility surfaces, enabling consistent pricing of the
derivatives (an example is given in \cite{dupire1997pah}).
Thirdly, the local volatility model is able to account for a
greater degree of empirical observations and theoretical arguments
on volatility than time dependent volatility (for instance the
leverage effect).

Some common local volatility models are: the Constant Elasticity
of Variance Model (CEV) proposed by Cox and Ross
\cite{cox1976ValCEV}, Mixture Distribution Models by Brigo and
Mercurio \cite{brigo2000mus} and the Implied Local Volatility by
Dupire \cite{DUPILoc}, Derman and Kani \cite{derman1994rs}.

\subsection{Stochastic Volatility Model: Definition and Characteristics}\label{Stochastic Volatility Model Definition and Characteristics section}
Although local volatility models were an improvement on time
dependent volatility, they possessed certain undesirable
properties. For
 example, volatility is perfectly correlated (positively or negatively) with
stock price %\cite{fouque2000dfm,p39}
yet empirical observations suggest no perfect correlation exists. % \cite{hobson1996sv}
 Stock prices empirically exhibit volatility clustering but under
local volatility this does not necessarily occur. Consequently
after local volatility development, models were proposed that
allowed volatility to be governed by its own stochastic process.
We now define stochastic volatility.
%\cite{fouque2000dfm,p40}:
\begin{defi}
Assume X follows the stochastic differential equation
\begin{eqnarray}
 dX/X &=& \mu dt + \sigma(\omega)dW_{1}.
%d\sigma_{1}/\sigma_{1} &=& \mu_{2}dt+ \sigma_{2}dW_{2}
\end{eqnarray}
Volatility is stochastic if $\sigma(\omega)$ is governed by a
stochastic process that is driven by another (but possibly correlated) random process, typically another Wiener process $dW_{2}$. % FOUQUE BOOK P41:
The probability space $(\Omega,\mathcal{F},\mathbb{P})$ is
$\Omega=\mathcal{C}([0,\infty):\mathbb{R}^{2})$, with filtration
$\{\mathcal{F}_{t}\}_{t\geq 0}$ representing information on two
Wiener processes $(W_{1},W_{2})$.\\
The process governing $\sigma(\omega)$ must always be positive for
all values since volatility can only be positive.
%%%%%%%%%%%%%%%%%%%%%%%%%%%%%%%%%%%%%%%%%%%%%%%%%%%%%%%%%%%%%%%%%%%%%%%%%%%%%%%%%%%%%%%%%%%%%%%%%%%%%%%%%%%%%%%%%%%%%%%%%%%%
\begin{comment}
%\cite{fouque2000dfm,p40}
\begin{itemize}
%  \item satisfy regularity conditions -LIPSHITZ, GROWTH,
 % OKSENDAL
  \item must always be positive for all values since volatility can only be
  positive;
  \item does not need to be an Ito process (e.g. jump processes and
  Markov chains are acceptable).
\end{itemize}
\end{comment}
%%%%%%%%%%%%%%%%%%%%%%%%%%%%%%%%%%%%%%%%%%%%%%%%%%%%%%%%%%%%%%%%%%%%%%%%%%%%%%%%%%%%%%%%%%%%%%%%%%%%%%%%%%%%%%%%%%%%%%%%%%%%%
The Wiener processes have instantaneous correlation $\rho$ $\in
[-1,1]$ defined by:
\[
corr(dW_{1}(t),dW_{2}(t))=\rho dt.
\]
\end{defi}% FOUQUE BOOK P41:
Empirically $\rho$ tends to be negative in equity markets due to
the leverage effect but close to 0 in
the currency markets. %\cite{FOUQDer}.
%Note that we can re-express:
%\begin{eqnarray}
%dW_{2} = \rho W_{1}+\sqrt{1-\rho^2}W_{2}
%\end{eqnarray}
Significant stochastic volatility models include: Johnson and
Shanno \cite{johnson1987opv},  Scott Model \cite{scott1987opv},
 Hull-White Model \cite{hull1987poa}, Stein and Stein  Model \cite{stein1991spd}, Heston model
\cite{heston1993cfs}

The key difference between local and stochastic volatility is that
local volatility  is not driven by a random process of its own;
there exists only one source of randomness ($dW_{1}$). In
stochastic volatility models, volatility has its own source of
randomness ($dW_{2}$) making volatility intrinsically stochastic.
We can therefore never definitely determine the volatility's
value, unlike in local volatility.
%Hence due to the introduction of another source of randomness,
%stochastic volatility models are 2 factor models.
%Thus the price SDE becomes:
%\begin{eqnarray*}
%dS(t)&=& \mu S(t)dt+\sigma(\omega,t) S(t)dB_{1}(t)\\
%\end{eqnarray*}

The disadvantages of stochastic volatility are firstly that these
models introduce a non-tradable source of randomness, hence the
market is no longer complete and we can no longer uniquely price
options or perfectly hedge. Therefore the practical applications
of stochastic volatility  are limited. Secondly stochastic
volatility models tend to be analytically less tractable. In fact,
it is common for stochastic volatility models to have no closed
form solutions for option prices. Consequently option prices can
only be calculated by simulation.

%\textbf{Stochastic Volatility and the Driving Process}\\
Stochastic volatility models fundamentally differ according to
their driving mechanisms for their volatility process. Different
driving mechanisms maybe favoured due to their tractability,
theoretical or empirical appeal and we can categorise stochastic
volatility models according to them.
%%%%%%%%%%%%%%%%%%%%%%%%%%%%%%%%%%%%%%%%%%%%%%%%%%%%%%%%%%%%%%%%%%%%%%%%%%%%%%%%%%%%%%%%%%%%%%%%%%%%%%%%%%%%%%%%%%
\begin{comment}
 Fouque (\cite{FOUQDer},p42) conveniently
lists the most popular driving processes:
\\\\
\begin{tabular}
[l]{|p{3cm}|p{3cm}|p{2cm}|p{4cm}|}\hline Model & Correlation
$\rho$ & $\sigma$=f(y) & y process \\\hline Hull-White & 0 &
f(y)=$\sqrt{y}$ & lognormal \\\hline Scott & 0 & f(y)=$e^{y}$ &
Mean-reverting OU
\\\hline Stein-Stein & 0 & f(y)=$\mid y \mid$ & Mean-reverting OU
\\\hline Ball-Roma & 0 & f(y)=$\sqrt{y}$ & CIR\\\hline Heston &
$\rho\neq0$  & f(y)=$\sqrt{y}$ & CIR
\\\hline
\end{tabular}
\\\\\\
\end{comment}
%%%%%%%%%%%%%%%%%%%%%%%%%%%%%%%%%%%%%%%%%%%%%%%%%%%%%%%%%%%%%%%%%%%%%%%%%%%%%%%%%%%%%%%%%%%%%%%%%%%%%%%%%%%%%%%%%%
Many stochastic volatility models favour a mean reverting driving
process. A mean reverting stochastic volatility process is of the
form \cite{fouque2000dfm}:%p40:
\begin{eqnarray}
 \sigma &=& f(Y),\\
 dY &=& \alpha(m-Y)dt+\beta dW_{2},\label{mean
 revert eqn definition}
 \end{eqnarray}
where:
\begin{itemize}
 \item $\beta \geq 0$ and $\beta$ is a constant;
  \item m is the long run mean value of $\sigma$;
  \item $\alpha$ is the rate of mean reversion.
\end{itemize}
Mean reversion is the tendency for a process to revert around its
long run mean value. We can economically account for the existence
of mean reversion through the cobweb theorem, which claims prices
mean revert due to lags in supply and demand \cite{lipsey1992fpe}.
The inclusion of mean reversion ($\alpha$) within volatility is
important, in particular, it controls the degree of
volatility clustering (burstiness) if all other parameters are unchanged. Volatility %p68,87
 clustering is an important empirical characteristic of
many economic or financial time series \cite{engle1982ach}, which
neither local nor time dependent volatility models necessarily
capture.
%\cite{HULLIntro}\cite{FOUQDer,p.59}
Additionally, a high $\frac{1}{\alpha}$ can be thought of as the
time required to decorrelate or ``forget"
its previous value.%p66,87

The equation (\ref{mean revert eqn definition}) is an
Ornstein-Uhlenbeck process in Y. It is known the solution to
equation
(\ref{mean revert eqn definition}) is:  %\cite{fouque2000dfm,p41}:
\begin{eqnarray}
Y(t) = m+(Y(0)-m)e^{-\alpha t}+\beta \int^{t}_{0}e^{-\alpha
t}dW_{2},
%
%removed s in \alpha(t-s) since i think just initial time from http://planetmath.org/encyclopedia/OrnsteinUhlenbeckProcess.html
%
%\sigma(t) = m+(\sigma(0)-m)e^{-\alpha t}+\beta
%\int^{t}_{0}e^{-\alpha(t-s)}dW_{2},
\end{eqnarray}
 where Y(t) has the
distribution
\begin{eqnarray}\label{OU dist eqn}
Y(t) \sim \mathcal{N} \left(m+(Y(0)-m)e^{-\alpha
t},\frac{\beta^2}{2\alpha}(1-e^{-2\alpha t})\right).
\end{eqnarray}
Note that alternative processes to equation (\ref{mean revert eqn
definition}) could have been proposed to define volatility as a
mean reverting stochastic volatility model, for example the Feller
or Cox-Ingersoll-Ross (CIR) process
\cite{fouque2000mrs}: %p8
\begin{eqnarray}
dY &=&  \alpha(m-Y)dt+\beta\sqrt{Y}dW_{2}.
\end{eqnarray}
However with $\sigma=f(Y)$ in equation (\ref{mean revert eqn
definition}) we can represent a broad range of mean reverting
stochastic volatility models in terms of a function of Y.
%%%%%%%%%%%%%%%%%%%%%%%%%%%%%%%%%%%%%%%%%%%%%%%%%%%%%%%%%%%%%%%%%%%%%%%%%%%%%%%%%%%%%%%%%%%%%%%%%%%%%%%%%%%%%%%%%%%%%%%%%%%%%%
\begin{comment}DON'T INCLUDE AS DON'T UNDERSTAND

%The OU can be geometric OU e.g. Wiggins model or  squared OU e.g.
%Stein and Stein model or inverse-gamma model and is the continuous
%time limit of the GARCH(1,1) model.
%NEED ADDRESS OF ISSUE OF WHY DIFFERENT PROCESSES CHOSEN -WHAT
%EMPIRICAL AND THEORETICAL EVIDENCE SUPPORT AND WHAT PROCESS MEAN.
\begin{itemize}
  \item
\item Feller or Cox-Ingersoll-Ross (CIR) process:
\begin{eqnarray}
d\sigma &=&  \alpha(m-\sigma)dt+\beta\sqrt{\sigma}dW_{2}.
\end{eqnarray}
\item Squared volatility process:
\begin{eqnarray}
d\sigma^{2} = \alpha(m-\sigma^{2})dt+\sigma \beta dW_{2}.
\end{eqnarray}
%\item the lognormal process: -NOT MEAN REVERTING, FOUQUE P40
%\begin{eqnarray*}
%d\sigma/\sigma &=& adt+bdW_{2}
%\end{eqnarray*}

\end{itemize}
\end{comment}
%%%%%%%%%%%%%%%%%%%%%%%%%%%%%%%%%%%%%%%%%%%%%%%%%%%%%%%%%%%%%%%%%%%%%%%%%%%%%%%%%%%%%%%%%%%%%%%%%%%%%%%%%%%%%%%%%%%%%%%%%%%%%%

\subsection{Regime Switching Volatility}
\subsubsection{General Regime Switching}
 A class of models that
address fundamental and long term volatility modelling is the
regime switching model (or hidden Markov model) e.g. as discussed
in \cite{timmermann2000mms},\cite{elliott1997ahm}. In fact,
Schwert suggests in \cite{schwert1989dsv} that volatility changes
during the Great Depression can be accounted
for by a regime change such as in Hamilton's regime switching model \cite{hamilton1989nae}.  %p1146
For regime switching models, generally the return distribution
rather than  the continuous time process is specified. A typical
example of a regime switching model is Hardy's model
\cite{hardy2001rsm}:
\begin{eqnarray}
log((X(t+1)/X(t))|i) \sim \mathcal{N}(u_{i},\varphi_{i}), i \in
\{1,..,R \}, \label{Hardy eqn}
\end{eqnarray}
where
\begin{itemize}
  \item $\varphi_{i}$ and $u_{i}$ are constant for the duration of the
regime;
    \item i denotes the current regime (also called the Markov
  state or hidden Markov state);
\item R denotes the total number of regimes;%, where $R \in \mathbb{N},R>0$.
\item a transition matrix \textbf{A} is specified.
\end{itemize}
For Hardy's model the regime changes discretely in monthly time
steps but stochastically, according to a Markov process.

Due to the ability of regime switching models to capture long term
and fundamental changes, regime switching models are primarily
focussed on modelling the long term behaviour, rather than the
continuous time dynamics. Therefore regime switching models switch
regimes over time periods of months, rather than switching in
continuous time. Examples of regime switching models that model
dynamics over shorter time periods are Valls-Pereira et
al.\cite{vallspereira2004pva}, who propose a regime switching
GARCH process, while Hamilton and Susmel \cite{hamilton1994ach}
give a regime switching ARCH process. Note that economic variables
other than stock returns, such as inflation, can also be modelled
using regime switching models.

%\label{Regime Switching Volatility Model Introduction}
%\subsection{Regime Switching Theory: Hidden Markov Models}\label{Regime
%Switching Volatility Theory section}
The theory of Markov models (MM) and Hidden Markov models (HMM)
are methods of mathematically modelling time varying dynamics of
certain statistical processes, requiring a weak set of assumptions
yet allow us to deduce a significant number of properties. MM and
HMM model a stochastic process (or any system) as a set of states
with each state possessing a set of signals or observations. The
models have been used in diverse applications such as  economics
\cite{sola2002tvs}, queuing theory \cite{salih2006maa},
engineering \cite{trentin2001sha} and biological modelling
\cite{melodelima2006cpi}. Following Taylor \cite{taylor1984ism} we
define a Markov model:
\begin{defi}
A Markov model is a stochastic process $X(t)$ with a countable set
of states and possesses the Markov property:
\begin{eqnarray}
p(q_{t+1}=j \mid q_{1},q_{2},..,q_{t}=i)=p(q_{t+1}=j \mid
q_{t}=i),
\end{eqnarray}
where
\begin{itemize}
  \item $q_{t}$ is the Markov state (or regime) at time t of
  $X(t)$;
  \item i and j are specific Markov states.
\end{itemize}
%$q_{1}$ to $q_{t-1}$ can be any states.
%That is, given $X_{t}$ is in state $i_{t}$ (denoted $X_{t}=i_{t}$), $X_{t+1}$ is not
%affected by the set of values $X_{u}$ for $u<t$.
\end{defi}
As time passes the process may remain or change to another state
(known as state transition). The state transition probability
matrix (also known as the \emph{transition kernel or stochastic
matrix}) $\mathbf{A}$, with elements $a_{ij}$, tells us the
probability of the process changing to state \emph{j} given that
we are now in state \emph{i}, that is $a_{ij}=p(q_{t+1}=j \mid
q_{t}=i)$. Note that $a_{ij}$ is subject to the standard
probability constraints:
\begin{eqnarray}
0 \leq a_{ij} & \leq &  1, \forall  i,j,\\
\sum^{\infty}_{j=1} a_{ij} &=& 1, \forall  i.
\end{eqnarray}
We assume that all probabilities are stationary in time. From the
definition of a MM the following proposition follows:
\begin{prop} % (see \cite{taylor1984ism}p97)
A  Markov model is completely defined once the following
parameters are known:
\begin{itemize}
  \item \emph{R}, the  total number of regimes or (hidden) states; % where $R \in \mathbb{N},R>0$;
\item state transition probability matrix $\mathbf{A}$ of size $R\times R$. Each element is
$a_{ij}=p(q_{t+1}=j|q_{t}=i)$, where i refers to the matrix row
number and j to the column number of $\mathbf{A}$;
  \item initial (t=1) state probabilities $\pi_{i}=p(q_{1}=i), \forall i$.
  \end{itemize}
\end{prop}
A hidden Markov model is simply a Markov model where we assume (as
a modeller) we do not observe the Markov states. Instead of
observing the Markov states (as in standard Markov models) we
detect observations or time series data where each observation is
assumed to be a function of the hidden Markov state, thus enabling
statistical inferences about the HMM.
%An observation type cannot be exclusive to just one state
%otherwise we know with certainty the current hidden state whenever
%that signal is emitted.
%The signals' probabilities % which must be a function of each state,
%are represented by a matrix .
Note that in a HMM it is the states which must be governed by a
Markov process, not the observations and throughout we will assume
one observation occurs after one state
transition. % -they are simply a probabilistic function of each state.
%\\\\
%Note that no observation matrix exists in MM since we can directly
%observe the states.
%Note: in MM/HMM it is NOT necessary for an output emission to be
%associated with a state transition event. In fact if we know which
%output emission solely gives state change then we can determine
%definitely which state we are in.

%\textbf{Hidden Markov Model Parameters}\\
\begin{prop}\label{HMM def prop}
A hidden Markov model is fully defined  when
the parameter set $\{\mathbf{A},\mathbf{B},\mathbf{\pi}\}$ are known: % \cite{rabiner1989thm}:
\begin{itemize}
  \item \emph{R}, the  total number of (hidden) states or regimes;
\item \textbf{A}, the (hidden) state transition matrix of size $R\times R$.
Each element is $a_{ij}=p(q_{t+1}=j|q_{t}=i)$;
  \item initial (t=1) state probabilities $\pi_{i}=p(q_{1}=i), \forall i$;
%\item $\{v_{1},v_{2} \ldots
%,v_{D}\}$, the set of observable discrete signals  where \emph{D}
%is the total number of observable signals.  Note that \emph{D} is
%always independent of \emph{N};
\item $\textbf{B}$, the observation matrix, where each entry is
$b_{j}(O_{t})=p(O_{t}|j)$ for observation $O_{t}$. For
$b_{j}(O_{t})$ is typically defined to follow some continuous
distribution e.g. $b_{j}(O_{t}) \sim
\mathcal{N}(u_{j},\varphi_{j})$.
%NOTE: don't need worry about v_{k} but needed to explain HMM, even though not used in GM HMM
%$\textbf{B}$, the observation matrix, where each entry is
%$b_{j}(d)=p(v_{d}|j)$, $d \in D$. The term $p(v_{d}|j)$ is the
%probability of observation $v_{d}$, given that we are in state $j$.
%The emission probs may be given but if not then they must be
%calculated/inferred.
%Note that a signal emission that causes no state transition simply
%means the arc for that signal transmission "re-loops" back to the same state.
\end{itemize}
%In our model, we assume observations v are modelled by a
%univariate mixture of Gaussians.
\end{prop}

Regime switching has been developed by various researchers. For
example, Kim and Yoo \cite{kim1995nic} develop a multivariate
regime switching model for coincident economic indicators. Honda
\cite{honda2003opc} determines the optimal portfolio choice in
terms of utility for assets following GBM but with  continuous
time regime switching
mean returns. %paper p49
Alexander and Kaeck \cite{alexander2007rdd} apply regime switching
to credit default swap spreads, Durland and McCurdy
\cite{durland1994ddt} propose a model with a transition matrix
that specifies state durations.

\subsubsection{Regime Switching Mean Reverting Stochastic Volatility Models}
%\subsubsection{Regime Switching Volatility Model Introduction}\label{Regime Switching Volatility Model Introduction}
A common shortfall in all the volatility models reviewed so far
has been that all these models are short \textit{or} long term
volatility models. The models implicitly ignore any long term or
broader economic factors influencing the volatility model, which
is empirically unrealistic and theoretically inconsistent.

Combining Wiener process driven stochastic volatility models with
regime switching would give the benefits of capturing the short
term price dynamics while the regime switching would capture
volatility changes due to fundamental and longer term effects in a
tractable method. In fact Alexander combines two models to model
short and long term dynamics, giving the binomial mixture
distribution model \cite{alexander2004nmd} (mixture distribution
model and a binomial tree).

A model that addresses both the long and short term dynamics are
regime switching mean reverting volatility models (RSMR). These
RSMR models are of the form where the intra-regime (within regime)
dynamics follow an exponential Ornstein-Uhlenbeck process (expOU):
\begin{eqnarray}
dX/X &=& \mu_{i}dt+ \sigma_{i}'(\omega)dW_{1},\\
 d\sigma_{i}'/\sigma_{i}' &=& \left[\alpha_{i}'(m_{i}'-ln\sigma_{i}') +  \dfrac{{\beta'}_{i}^{2}}{2} \right]dt+
 \beta_{i}'dW_{2},\\ %CORRECT TAKEN FROM SCOTT MODEL PAPER
% [\frac{\beta_{R}^{2}}{2} -\alpha_{R}(ln\sigma_{R}-m_{R})]dt+ \beta_{R}dW_{2}\\ %\Buchbinder paper
% exp\{\alpha_{R}(m_{R}-\sigma_{R})dt+\beta_{R}dW_{2} \}
% NOT SURE BELOW CORRECT
%
%\mbox{where }\sigma_{R}(t) &=& exp\{m_{R}+(\sigma_{R}(0)-m_{R})e^{-\alpha_{R} t}+\beta_{R}
%\int^{t}_{0}e^{-\alpha_{R}(t-s)}dW_{2}\},\\
i &\in & \{1,2\}.
\end{eqnarray}
%BELOW TRYING TO GIVE PRECISE SPEC OF MODEL AS POSSIBLE
Alternatively, it can be more convenient to express RSexpOU in
terms of OU:
\begin{eqnarray}\label{RSexpOU eqn}
dX/X &=& \mu_{i}dt+ \sigma_{i}(\omega)dW_{1},\\
\sigma_{i} &=& exp(Y_{i}(t)), \\
dY_{i}(t) &=&\alpha_{i}(m_{i}-Y_{i})dt+\beta_{i}dW_{2}.
\end{eqnarray}
 We denote i as the current regime and for convenience set R=2. The regime switching process is a discrete time Markov
process (e.g. changing in time intervals of one year).% Since regime changes occur instantaneously,  %no "transitional phase" exists in time
%the expOU process is always governed by one of the regimes.
We note that regime switching models that change state
continuously do not capture long term and sustained changes. In
fact Hamilton's regime switching models have discrete time
switching periods of three months (see for instance
\cite{engel1990lsd}).

The literature on regime switching models with stochastic
differential equations in their continuous time price dynamics is
currently growing but is not significant, especially for mean
reverting stochastic volatility. Examples include Kalimipalli
\cite{kalimipalli2004rss} and Smith \cite{smith2002msa}, who
describe a stochastic volatility model with regime switching to
model interest rates, Elliot et al. \cite{elliott2005opa} propose
a regime switching Geometric Brownian Motion model (GBM) and Mari
\cite{mari2006rsc} models electricity spot prices with each regime
possessing a different stochastic differential equation structure,
such as mean reversion in the drift.
%DON'T INCLUDE SIU -PROVIDES OPTION PRICES FOR RS-SV WITH SV=CIR.

\section{Perturbation Based Option Pricing}
We price options for RSMR model using Fouque's option pricing
method, which is based on perturbation theory. Fouque's
perturbation method is normally applied to mean reverting
stochastic volatility models; we briefly introduce Fouque's
perturbation theory by first introducing perturbation analysis and
then Fouque's method. Finally we show we can apply Fouque's method
to RSMR models. For more information on perturbation theory the
reader is referred to Hinch \cite{hinch1991pm}, Simmonds
\cite{simmonds1998flp}, Bender and Orszag \cite{bender1999amm}.

\subsection{Introduction to Perturbation Analysis}
For many mathematical equations we cannot find precise solutions,
 however, we would like to obtain approximations in a mathematically
consistent and rigorous way. The perturbation method is applied
when we know the solution to an equation  but would like to know
the solutions when variables are minutely changed (or perturbed)
by a small amount $\epsilon$, where $\epsilon \rightarrow 0$ but
is finite. Perturbation problems can be solved by an iterative
method or by a series expansion method;
Fouque adopts the latter method.%\cite{hinch1991pm}p1-4

Perturbation analysis can be explained by example;  consider the
following example equation we wish to solve:
\begin{eqnarray}
x^{2}-1=x.
\end{eqnarray}
This equation can be easily solved using knowledge of quadratics.
Now consider when we perturb this problem by $\epsilon$:
\begin{eqnarray}\label{perturbed eqn}
x^{2}-1=\epsilon x,\mbox{where }\epsilon \rightarrow 0.
\end{eqnarray}
 Solving
for x now becomes difficult for various small (but finite) values
of $\epsilon$. Note that in this example it is still possible to
find the 2 analytic solutions to equation (\ref{perturbed eqn})
using our knowledge of quadratics:
\begin{eqnarray}
x_{a} &=& \epsilon/2+\sqrt{(1+\epsilon^{2})/4},\\
x_{b} &=& \epsilon/2-\sqrt{(1+\epsilon^{2})/4}.
\end{eqnarray}
However, instead of using knowledge of quadratics we apply the
perturbation method: we determine the solutions $x_{a},x_{b}$ by
\textit{approximating} x by a series expansion. One could use any
series expansion (e.g. Taylor, Maclaurin) but we use a power
series expansion:
\begin{eqnarray}
x &=& \sum_{n=0}^{\infty}x_{n}\epsilon^{n}\\ \label{series approx
eqn}
   &=& x_{0}+\epsilon
x_{1}+\epsilon^{2}x_{2}+\epsilon^{3}x_{3}+....
\end{eqnarray}
where $x_{1},x_{2},x_{3}$ are constants to be determined. Note
that $x_{0}$ is called the leading term, $x_{1}$ the first
correction, $x_{2}$ the second correction and so on. The
unperturbed or known solution corresponds to $\epsilon=0$ and
therefore equals $x_{0}$. Normally obtaining a solution to the
unperturbed problem (that is $x_{0}$) is easy to obtain, otherwise
the set of equations (as shown later) that need to be solved to
obtain approximations to $x_{a},x_{b}$ will be difficult to solve.

With the series approximation of x (equation (\ref{series approx
eqn})), we insert this approximation into the original equation
(\ref{perturbed eqn}):
\begin{eqnarray*}
x^{2}-1 &=& \epsilon x\\
(x_{0}+\epsilon x_{1}+\epsilon^{2}x_{2}+\epsilon^{3}x_{3})^{2}-1 &=& \epsilon(x_{0}+\epsilon x_{1}+\epsilon^{2}x_{2}+\epsilon^{3}x_{3})\\
(x_{0}^{2}+2\epsilon
x_{0}x_{1}+\epsilon^{2}(x_{1}^{2}+2x_{0}x_{2})+\mathcal{O}(\epsilon^{3})+\mathcal{O}(\epsilon^{4}))-1
&=& \epsilon(x_{0}+\epsilon x_{1}+\mathcal{O}(\epsilon^{3})).
\end{eqnarray*}
Now it is standard practice to ignore terms of order
$\epsilon^{3}$ (denoted $\mathcal{O}(\epsilon^{3})$) and higher
since $\epsilon$ is small and so terms of
$\mathcal{O}(\epsilon^{3})$ or higher are considered
insignificant. Therefore the equation simplifies to
\begin{eqnarray}
x_{0}^{2}+2\epsilon
x_{0}x_{1}+\epsilon^{2}(x_{1}^{2}+2x_{0}x_{2})-1 &=& \epsilon
x_{0}+\epsilon^{2} x_{1}\\ \label{final line pert eqn}
(x_{0}^{2}-1)+ \epsilon(2x_{0}x_{1}-x_{0}) + \epsilon^{2}(
x_{1}^{2} + 2x_{0}x_{2}- x_{1}) &=& 0.
\end{eqnarray}
Now equation (\ref{final line pert eqn}) is significant because
the bracketed terms represent coefficients of  $\epsilon$ for each
power and each bracketed term separately must equal 0 for the
equation to balance. So now we can equate terms of the same order
from equation (\ref{final line pert eqn}):
\begin{eqnarray}\label{pert order 0 eqn}
\mathcal{O}(\epsilon^{0})&:& x_{0}^{2}-1 = 0,\\
\mathcal{O}(\epsilon^{1})&:& 2x_{0}x_{1}-x_{0} = 0,\\\label{pert
order 2 eqn} \mathcal{O}(\epsilon^{2})&:& x_{1}^{2} + 2x_{0}x_{2}-
x_{1} = 0.
\end{eqnarray}
Solving the equations (\ref{pert order 0 eqn})-(\ref{pert order 2
eqn}) and remembering there are 2 solutions gives:
\begin{eqnarray}
x_{a}&:&\{x_{0} = 1,x_{1}=1/2, x_{2}=1/8\}, \\
x_{b}&:&\{x_{0} = -1,x_{1}=1/2, x_{2}=-1/8\}.
\end{eqnarray}
Using these values, we can now calculate approximate but extremely
accurate solutions to $x_{a},x_{b}$:
\begin{eqnarray}
x & = & x_{0}+\epsilon x_{1}+\epsilon^{2}x_{2}+....\\
x_{a} & = & 1+\epsilon(1/2)+ \epsilon^{2}(1/8)\\
x_{b} & = & -1+\epsilon(1/2)+ \epsilon^{2}(-1/8).
\end{eqnarray}

To summarise, the three steps of perturbative analysis (by Bender
and
Orszag \cite{bender1999amm}) are: %p320
firstly to convert the original (unperturbed) problem into a
  perturbed problem by introducing $\epsilon$. Secondly, we assume
  the expression for the solution is in the form of a
  perturbation series and determine the coefficients of that
  series. Thirdly, we determine answers to the solutions by summing the perturbation
  series for the appropriate values of $\epsilon$.

Perturbation problems can be divided into two categories: regular and %\cite{simmonds1998flp}p1
singular perturbation problems. Singular pertubation problems have
significantly different solutions as $\epsilon \rightarrow 0$ when
compared to the unperturbed problem ($\epsilon=0$). If the
solution changes insignificantly with $\epsilon$ then we have a
regular perturbation;  the perturbed solution smoothly approaches
the unperturbed solution as $\epsilon \rightarrow 0$. Fouque's
perturbation based option pricing method is a singular perturbation problem. %\cite{bender1999amm}p324,\cite{hinch1991pm}p

To give an example of perturbation problems, a spinning drawing
pin rotates around its vertical axis and over a short time period
its horizontal distance from its vertical axis is unlikely to be
significantly different from its initial position (at t=0).
However over long time periods, the drawing pin will eventually
come to a physical rest and its horizontal position will be
appreciably different than from at t=0. Therefore over a small
time period the position can be modelled as a regular perturbation
problem but over long time periods it is a singular perturbation
problem.

\subsection{Perturbation Formulation of Option Pricing}\label{Perturbation Formulation of Stochastic Volatility
Option Pricing}
 Option pricing where the stock price process is
governed by stochastic volatility can be analytically challenging.
Fouque \cite{fouque2000dfm} finds a method of pricing options
whose underlying is governed by a mean reverting stochastic
volatility process by solving the option pricing partial
differential equation as a
singular perturbation problem.  A  %\cite{fouque2000dfm,p90}
significant benefit of Fouque's method is that the option pricing
method is model independent;  it can be applied to any fast mean
reverting stochastic volatility model and without explicit
knowledge of the volatility dynamics. Additionally, the option
pricing calibration is far more parsimonious compared to other
stochastic volatility option pricing methods.

To find the option pricing partial differential equation  when the
underlying asset is driven by mean reverting stochastic
volatility, we can use a replicating portfolio argument.
Alternatively, we can determine the option pricing partial
differential equation associated with an option by risk neutral
valuation and apply the Feynman-Kac equation, which is the
approach we take here.  We abbreviate call option
C(X,K,s,T,r,f(Y)) to C(s,X,Y), which is an option at time s with
volatility as a function of some mean reverting stochastic
volatility process (f(Y)). By risk neutral valuation C(s,X,Y) is
given by:
\begin{eqnarray}
C(s,X,Y)=e^{-r(T-s)}E^{\mathbb{Q}}[X(T)-K|\mathcal{F}_{s}]^{+}.
\end{eqnarray}
We recall that in a no arbitrage market under stochastic
volatility there is more than one risk neutral measure
$\mathbb{Q}$ (no unique $\mathbb{Q}$ measure exists). The possible
$\mathbb{Q}$ measures can be parameterised by the market price of
volatility risk $\gamma=\gamma(t,X,f(Y))$. Therefore $\mathbb{Q}$
measures
are a function of $\gamma$, %where as in Fouque\cite{fouque2000dfm}
consequently the price of option C(t,X,Y) is a function of $\gamma$.%following armerin paper

%p44-6 argue \gamma=\gamma(t,X,Y) by replicating portfolio argument and saying function must have this parameterisation by deduction
%Note that in Fouque's method we do not directly determine $\gamma$
%but extract it from empirical option data.

To obtain the (risk neutral) option pricing partial differential
equation by the Feynman-Kac equation, we require the risk neutral
process. We begin with a stock price with a mean reverting
stochastic volatility process, that is:
\begin{eqnarray}
dX/X &=& \mu dt + \sigma dW_{1},\\
\sigma &=& f(Y),\\
 dY &=& \alpha(m-Y)dt+\beta dW_{2}.
\end{eqnarray}
The risk neutral process for dX/X is:
\begin{eqnarray}
dX/X &=& rdt + \sigma dW_{1}^{\mathbb{Q}}.
\end{eqnarray}
%To obtain the risk neutral process for stochastic volatility we
%follow Fouque's method \cite{fouque2000dfm}chapter 2.%p41
We can re-express $dW_{2}$ as \cite{fouque2000dfm}:
\begin{eqnarray}\label{dW2 eqn}
dW_{2}=\rho dW_{1}+\sqrt{1-\rho^{2}}dW^{*},
\end{eqnarray}
where $W^{*}$ is a standard Wiener process independent of
$W_{1}$.  %working with %2 independent Wiener processes ($W_{1},W^{*}$) facilitates derivation compared to ($W_{1},W_{2}$).
%applying the multidimensional Girsanov's Theorem rather than to ($W_{1},W_{2}$).%\cite{armerin:sv}p13
To change dY to its risk neutral process we require:
\begin{enumerate}
\item
\begin{eqnarray}
dW_{1}^{\mathbb{Q}} &=& dW_{1}+\dfrac{\mu-r}{f(Y)}dt,\\
\label{dW1Q eqn} \mbox{i.e. }dW_{1} &=&
dW_{1}^{\mathbb{Q}}-\dfrac{\mu-r}{f(Y)}dt.
\end{eqnarray}
\item
\begin{eqnarray}
dW^{\mathbb{Q}*} &=& dW^{*}+\gamma(t,X,Y) dt,\\ \label{dW*Q eqn}
\mbox{i.e. }dW^{*}&=& dW^{\mathbb{Q}*}-\gamma(t,X,Y) dt.
\end{eqnarray}
\end{enumerate}
To obtain the risk neutral process of dY we substitute in
equations (\ref{dW2 eqn})-(\ref{dW*Q eqn}) and rearrange dY;
starting with substituting equation (\ref{dW2 eqn}) for $dW_{2}$
gives
\begin{eqnarray*}
dY &=&  \alpha(m-Y)dt+\beta dW_{2},\\
 &=& \alpha(m-Y)dt+\beta \left[\rho
       dW_{1}+\sqrt{1-\rho^{2}}dW^{*} \right].
\end{eqnarray*}
Substituting  equation (\ref{dW1Q eqn}) for $dW_{1}$ and equation
(\ref{dW*Q eqn}) for $dW^{*}$ gives:
\begin{eqnarray*}
dY &=& \alpha(m-Y)dt\\
 &+& \beta \left[ \rho
       \left(dW_{1}^{\mathbb{Q}}-\dfrac{\mu-r}{f(Y)}dt \right)
       + (\sqrt{1-\rho^{2}})(dW^{\mathbb{Q}*}-\gamma(t,X,Y)
       dt) \right].
\end{eqnarray*}
Rearranging now gives:
\begin{eqnarray*}
dY      &=&
       \left[\alpha(m-Y)-\beta \left(\rho\dfrac{(\mu-r)}{f(Y)}+\sqrt{1-\rho^{2}}\gamma(t,X,Y) \right) \right]dt\\
       &+& \beta \left[\rho dW_{1}^{\mathbb{Q}}+\sqrt{1-\rho^{2}}dW^{\mathbb{Q}*} \right].
%       &=& \left[\alpha(m-Y)-\beta \left(\rho\dfrac{(\mu-r)}{f(Y)}+\sqrt{1-\rho^{2}}\gamma(t,X,Y) \right) \right]dt\\
%       &+& \beta dW_{2}^{Q},\\
%\mbox{where }dW_{2}^{Q} = \rho dW_{1}^{Q} &+&
%\sqrt{1-\rho^{2}}dW^{Q*}.
\end{eqnarray*}

Since the risk neutral equation for dX is a two dimensional
diffusion equation in $(W_{1}^{\mathbb{Q}},W^{\mathbb{Q}*})$, the
associated option pricing  partial differential equation for
C(s,X,Y) is obtained by the application of the multidimensional
version of the Feynman-Kac equation with boundary condition
$C(T,x,y)$.
\\\\
The Feyman-Kac equations give the option pricing partial differential equation:% Fouque p45
 \begin{eqnarray*}
0 &=& \dfrac{\partial C}{\partial s} +
\dfrac{1}{2}f^{2}(y)x^{2}\dfrac{\partial^{2} C}{\partial x^{2}} +
\rho \beta xf(y)\dfrac{\partial^{2} C}{\partial x
\partial y} +\dfrac{\beta^{2}}{2}\dfrac{\partial^{2}
C}{\partial y^{2}}+ r\left(x\dfrac{\partial C}{\partial x} -C \right)\\
&+& [\alpha(m-y)-\beta\Lambda(s,x,y)]\dfrac{\partial C}{\partial
y},\\
\mbox{where }\Lambda(s,x,y) &=&
\rho\dfrac{(\mu-r)}{f(y)}+\sqrt{1-\rho^{2}}\gamma(s,x,y).
\end{eqnarray*}
The term $\Lambda(t,x,y)$ can be interpretted as a  combination of
market price of risk and market price of volatility risk. Fouque
assumes that $\Lambda(s,x,y)=\Lambda(y)$, a function of y only: %fouque p88,
\begin{eqnarray}
\Lambda(y) &=&
\rho\dfrac{(\mu-r)}{f(y)}+\sqrt{1-\rho^{2}}\gamma(y).
\end{eqnarray}

Now from our knowledge of OU processes (which is also Y(t)) we
know as $t\rightarrow\infty$  the distribution becomes
$\mathcal{N}(m,\frac{\beta^2}{2\alpha})$ (see equation (\ref{OU
dist eqn})). If we denote the variance of this long term
distribution as $v^{2}$, that is
$v^{2}=\frac{\beta^{2}}{2\alpha}$, then $\beta=v\sqrt{2\alpha}$
and we can re-express the option pricing partial differential
equation as:
 \begin{eqnarray*}
0 &=& \dfrac{\partial C}{\partial s} +
\dfrac{1}{2}f^{2}(y)x^{2}\dfrac{\partial^{2} C}{\partial x^{2}} +
\rho v\sqrt{2\alpha} xf(y)\dfrac{\partial^{2} C}{\partial x
\partial y} +v^{2}\alpha\dfrac{\partial^{2}
C}{\partial y^{2}}+ r\left(x\dfrac{\partial C}{\partial x} -C \right)\\
&+& [\alpha(m-y)-v\sqrt{2\alpha}\Lambda(y)]\dfrac{\partial
C}{\partial y}.
\end{eqnarray*}

Now if we re-express $\alpha$ as $\alpha=\dfrac{1}{\epsilon}$,
where $\epsilon\rightarrow 0$ is small but finite (since
$\alpha\rightarrow \infty$) our equation becomes: %same as Fouque p89
\begin{eqnarray*}
0 &=& \dfrac{\partial C}{\partial s} +
\dfrac{1}{2}f^{2}(y)x^{2}\dfrac{\partial^{2} C}{\partial x^{2}} +
\rho \dfrac{v\sqrt{2}}{\sqrt{\epsilon}} xf(y)\dfrac{\partial^{2}
C}{\partial x
\partial y} +\dfrac{v^{2}}{\epsilon}\dfrac{\partial^{2}
C}{\partial y^{2}}+ r\left(x\dfrac{\partial C}{\partial x} -C \right)\\
&+&
\left[\dfrac{(m-y)}{\epsilon}-\dfrac{v\sqrt{2}}{\sqrt{\epsilon}}\Lambda(y)
\right]\dfrac{\partial C}{\partial y}.
\end{eqnarray*}

We can also re-express the equation using the following operator
notation for convenience:
%%%%%%%%%%%%%%%%%%%%%%%%%%%%%%%%%%%%%%%%%%%%%%%%%%%%%%%%%%%%%%%%%%%%%%%%%%%%%%%%%%%%%%%%%%%%%%%%%%%%%%%%%%%%%%%%%
\begin{comment}
LATEX FOR INDICATOR FN.
$$
1{\hskip -2.5 pt}\hbox{I} \qquad \qquad
$$
FROM http://www.statslab.cam.ac.uk/~elie/ext/tex/
\end{comment}
%%%%%%%%%%%%%%%%%%%%%%%%%%%%%%%%%%%%%%%%%%%%%%%%%%%%%%%%%%%%%%%%%%%%%%%%%%%%%%%%%%%%%%%%%%%%%%%%%%%%%%%%%%%%%%%%%
\begin{eqnarray}\label{LO operator definition}
%\mathcal{L}_{0} &=&
%\frac{\beta^{2}}{2}\frac{\partial^{2}}{\partial Y^{2}}+
%M(t,Y)\frac{\partial^{2}}{\partial Y}\\
\mathcal{L}_{0} &=& v^{2}\frac{\partial^{2}}{\partial y^{2}}+
(m-y)\frac{\partial}{\partial y}, \\ \label{L1 operator
definition} \mathcal{L}_{1} &=& \sqrt{2}\rho
vxf(y)\frac{\partial^{2}}{\partial x\partial
y}-\sqrt{2}v\Lambda(y)\frac{\partial}{\partial y},\\
\mathcal{L}_{2} &=& \frac{\partial}{\partial s} +
\frac{1}{2}f^{2}(y)x^{2}\frac{\partial^{2}}{\partial x^{2}}+
r\left(x\frac{\partial }{\partial x} -1{\hskip -2.5 pt}\hbox{I}
\right).
\end{eqnarray}
Therefore the option pricing partial differential equation is:
\begin{eqnarray}\label{MRSV option price eqn pert}
\left(\frac{1}{\epsilon}\mathcal{L}_{0}+
\frac{1}{\sqrt{\epsilon}}\mathcal{L}_{1}+
\mathcal{L}_{2}\right)C=0.
\end{eqnarray}
Equation (\ref{MRSV option price eqn pert}) is a (singular) perturbation problem;  %\cite{fouque2000dfm}p89
we can therefore solve the option pricing partial differential
equation (under mean reverting stochastic volatility) by
perturbation methods. We note that:
\begin{itemize}
  \item  we sometimes express
$\mathcal{L}_{2}=\mathcal{L}_{BS}(f(y))$, where
$\mathcal{L}_{BS}(\cdot)$ is the Black-Scholes partial
differential operator with volatility $(\cdot)$.

\item  the $\mathcal{L}_{0}$ operator gives a Poisson equation for Y (an OU process). Note
 that the popular definition of a Poisson equation is a
 form of Laplace equation (second order partial differential equation),
 however, for stochastic processes it is $\mathcal{L}_{0}$.
%%%%%%%%%%%%%%%%%%%%%%%%%%%%%%%%%%%%%%%%%%%%%%%%%%%%%%%%%%%%%%%%%%%%%%%%%%%%%%%%%%%%%%%%%%%%%%%%%%%%%%%%%%%%%%%%
\begin{comment}
 the infinitessimal
 operator. An infinitessimal generator $\mathcal{A}(g(x))$ on a Markov process X(t), with function g, X(0)=x, is defined as \cite{fouque2000dfm}:%p24
\begin{eqnarray}
\mathcal{A}(g(x))=\lim_{t\rightarrow 0}
\dfrac{E[g(X(t))]-g(x)}{t}.
\end{eqnarray}
\end{comment}
%%%%%%%%%%%%%%%%%%%%%%%%%%%%%%%%%%%%%%%%%%%%%%%%%%%%%%%%%%%%%%%%%%%%%%%%%%%%%%%%%%%%%%%%%%%%%%%%%%%%%%%%%%%%%%%%
\end{itemize}
%%%%%%%%%%%%%%%%%%%%%%%%%%%%%%%%%%%%%%%%%%%%%%%%%%%%%%%%%%%%%%%%%%%%%%%%%%%%%%%%%%%%%%%%%%%%%%%%%%%%%%%%%%%%%%%%
\begin{comment}
$\mathcal{L}_{1}$ contains the correlation $\rho$ and takes into
account correlation in stochastic
volatility. %It is the only operator containing $\rho$ and the operator disappears when $\rho$=0. -EQN NOT SAY THIS???
\end{comment}
%%%%%%%%%%%%%%%%%%%%%%%%%%%%%%%%%%%%%%%%%%%%%%%%%%%%%%%%%%%%%%%%%%%%%%%%%%%%%%%%%%%%%%%%%%%%%%%%%%%%%%%%%%%%%%%%

\subsection{Perturbation Solution to Option Pricing}\label{Perturbation Solution to Option Pricing section}
To determine the solution to C(s,x,y) by a perturbation approach
we expand C in powers of $\sqrt{\epsilon}$
\begin{eqnarray}\label{C pert exp eqn}
C=C_{0}+\sqrt{\epsilon}C_{1}+ \epsilon
C_{2}+\epsilon^{\frac{3}{2}} C_{3}+ ......
\end{eqnarray}
 and insert equation (\ref{C pert exp eqn}) into equation (\ref{MRSV option price eqn pert}):
\begin{eqnarray*}
0 &=& \left(\frac{1}{\epsilon}\mathcal{L}_{0}+
\frac{1}{\sqrt{\epsilon}}\mathcal{L}_{1}+
\mathcal{L}_{2}\right)(C_{0}+\sqrt{\epsilon}C_{1}+ \epsilon
C_{2}+\epsilon^{\frac{3}{2}} C_{3}+ ......)\\
& = & \frac{1}{\epsilon}\mathcal{L}_{0}C_{0} \\ \label{order 0} &
+ &
\frac{1}{\sqrt{\epsilon}}(\mathcal{L}_{0}C_{1}+\mathcal{L}_{1}C_{0})\\
\label{order 1}
 & + &
(\mathcal{L}_{0}C_{2}+\mathcal{L}_{1}C_{1}+\mathcal{L}_{2}C_{0})\\
\label{order 2}
 &+&
\sqrt{\epsilon}(\mathcal{L}_{0}C_{3}+\mathcal{L}_{1}C_{2}+\mathcal{L}_{2}C_{1})\\
&+& \mathcal{O}(\epsilon).
\end{eqnarray*}

As with standard perturbation problems, we can equate terms of the
same order to assist us in solving the equation. Fouque solves C
upto $C=C_{0}+\sqrt{\epsilon}C_{1}$ (although we could solve
further), therefore our aim is to find $C_{0}$ and $C_{1}$.
\\\\
\textbf{Zero Order Term ($C_{0}$)}\\
We begin by firstly determining $C_{0}$;  by equating terms of
order $1/\epsilon$ we have:
\begin{eqnarray}\label{C0 eqn}
\mathcal{L}_{0}C_{0}=0.
\end{eqnarray}
For equation (\ref{C0 eqn}) to be correct and since
$\mathcal{L}_{0}$ acts only on y, $C_{0}$ must be a function of
s,x only: $C_{0}=C_{0}(s,x)$.

Now equating terms of order $1/\sqrt{\epsilon}$ we have:
\begin{eqnarray}\label{C1 eqn}
\mathcal{L}_{0}C_{1}+\mathcal{L}_{1}C_{0}=0.
\end{eqnarray}
 For equation (\ref{C1 eqn}) to
be correct, if $\mathcal{L}_{1}$ takes derivatives on y only and
since we have already deduced $C_{0}=C_{0}(s,x)$ therefore we have
$\mathcal{L}_{1}C_{0}=0$. This in turn implies from equation
(\ref{C1 eqn}) that $\mathcal{L}_{0}C_{1}=0$, which in turn
implies $C_{1}=C_{1}(s,x)$ for equation (\ref{C1 eqn}) to be
correct. If desired one could continue to apply this iterative
method for higher orders 1, $\epsilon$ and so on to obtain further
equations.

Now the order 1 terms give:
\begin{eqnarray}
\mathcal{L}_{0}C_{2}+\mathcal{L}_{1}C_{1}+\mathcal{L}_{2}C_{0}=0.
\end{eqnarray}
Since we have already reasoned $\mathcal{L}_{1}C_{1}$=0 we can
write
\begin{eqnarray}\label{L0C2}
\mathcal{L}_{0}C_{2}+\mathcal{L}_{2}C_{0}=0.
\end{eqnarray} %Fouque p91
Regarding the variable x as fixed $\mathcal{L}_{2}C_{0}$ is a
function of y since $\mathcal{L}_{2}$ contains f(y);  if we focus
on the y dependency only we can rewrite equation (\ref{L0C2}) as:
\begin{eqnarray}\label{Poisson eqn1}
\mathcal{L}_{0}\chi(y)+g(y)=0,
\end{eqnarray}
where $g(y)=\mathcal{L}_{2}C_{0}$. Fouque states this is a Poisson
equation in  $\chi(y)$ and so can assert for a solution to exist  %with respect to the operator $\mathcal{L}_{0}$ in the variable y.
%a Poisson equation is a second order partial differential equation that arises frequently in Physics.
g(y) must fulfil the ``centering condition":
\begin{eqnarray}\label{centering eqn}
<g(y)>=\int_{-\infty}^{\infty} g(y)\Phi(y)dy=0.%\\
%\mbox{where }<g(Y)>=\lim_{t\rightarrow \infty}E[g(Y)|Y_{0}]
\end{eqnarray}%must be p(y) not cdf because want e[x]
The terms in equation (\ref{centering eqn}) relate to the
theoretical area of ergodicity and will be covered in more detail
in section \ref{Regime Switching and Pertubation Based Option
Pricing}; we briefly define them for now. We define $<g(X)>$ as:
\begin{eqnarray}
<g(X)>=\lim_{t\rightarrow \infty} \dfrac{1}{t} \int_{0}^{t}g(X)ds.
\end{eqnarray}
For any process X that is ergodic or a function g() of an ergodic
process g(X), there exists an invariant distribution or long term
distribution exhibited by g(X) as time tends to a large limit. We
denote $\Phi(\cdot)$ as the probability measure of the invariant
distribution. Therefore
\begin{eqnarray}
E[g(X)]= \int_{-\infty}^{\infty} g(X)\Phi(X)dX.
\end{eqnarray}
Equation (\ref{centering eqn}) therefore means $<g(y)>$ equals its
average under its invariant distribution;  note that both $<g(y)>$
and $E^{\Phi}[g(X)]$ are constants. We also denote
\begin{eqnarray}
(\overline{\sigma})^{2} &=& <f^{2}(Y)>
\end{eqnarray}
when $f(Y)=\sigma$, that is when we operate $< >$ upon the squared
mean reverting stochastic volatility process. The variable
$\overline{\sigma}$ is a constant
and is referred to as the ``effective volatility" by Fouque \cite{fouque2000dfm}. %p87

From equation (\ref{Poisson eqn1}) $\mathcal{L}_{0}\chi=-g(y)$
therefore substituting this into equation (\ref{centering eqn})
gives:
\begin{eqnarray}
<\mathcal{L}_{2}C_{0}> &=& 0.
\end{eqnarray}
Since we have already established $C_{0}$ does not depend on y we
have
\begin{eqnarray}
<\mathcal{L}_{2}>C_{0} &=& 0.
\end{eqnarray}
Now $\overline{\sigma}^{2}=<f^{2}(y)>$ and so by definition of
$\mathcal{L}_{BS}$ we have
\begin{eqnarray}
<\mathcal{L}_{2}> &=& \mathcal{L}_{BS}(\overline{\sigma}),\\
\label{LBSC0=0}
 \mbox{so that } \mathcal{L}_{BS}(\overline{\sigma})C_{0}
&=& 0.
\end{eqnarray}
Therefore we finally deduce the value of $C_{0}$: it is the option
price obtained by the Black-Scholes equation with constant
volatility $\sigma=\overline{\sigma}$.
\\\\
\textbf{First Order Term ($C_{1}$)}\\
Since the centering condition $<\mathcal{L}_{2}C_{0}>$=0 is
satisfied we can re-express $\mathcal{L}_{2}C_{0}$. Recalling that
we can denote $\mathcal{L}_{2}C_{0}=\mathcal{L}_{BS}(f(y))C_{0}$
we have:
\begin{eqnarray}
\mathcal{L}_{2}C_{0} &=& \mathcal{L}_{2}C_{0}-<\mathcal{L}_{2}C_{0}>,\mbox{ since }<\mathcal{L}_{2}C_{0}>=0,\\
           &=&
           \mathcal{L}_{BS}(f(y))C_{0}-\mathcal{L}_{BS}(\overline{\sigma})C_{0}\\
&=&
\dfrac{1}{2}(f^{2}(y)-\overline{\sigma}^{2})x^{2}\dfrac{\partial^{2}C_{0}}{\partial
           x^{2}}.\label{L2C0}
\end{eqnarray}
%comment: last line possible by subtracting: {L}_{2}C_{0}-{L}_{BS}(\overline{\sigma})C_{0} -look at the operators
% only difference between 2 gives final line
From equation (\ref{L0C2}) we have:
\begin{eqnarray}
\mathcal{L}_{0}C_{2}+\mathcal{L}_{2}C_{0} &=& 0\\
\mathcal{L}_{0}C_{2} &=& -\mathcal{L}_{2}C_{0}.
\end{eqnarray}
Now substituting $\mathcal{L}_{2}C_{0}$ expression from  equation
(\ref{L2C0}) we have an expression for $C_{2}$:
\begin{eqnarray}
\mathcal{L}_{0}C_{2} &=& -\mathcal{L}_{2}C_{0}\\
           &=& -\left(\dfrac{1}{2}(f^{2}(y)-\overline{\sigma}^{2})x^{2}\dfrac{\partial^{2}C_{0}}{\partial
           x^{2}}\right)\\
\mbox{so that } C_{2} &=&
-\mathcal{L}_{0}^{-1}\left(\dfrac{1}{2}(f^{2}(y)-\overline{\sigma}^{2})x^{2}\dfrac{\partial^{2}C_{0}}{\partial
           x^{2}} \right)\\ \label{C2}
&=&
-\dfrac{1}{2}\mathcal{L}_{0}^{-1}(f^{2}(y)-\overline{\sigma}^{2})x^{2}\dfrac{\partial^{2}C_{0}}{\partial
x^{2}}.
\end{eqnarray}
The last line was possible since ${\mathcal{L}}_{0}$ takes
derivatives in y only. If we write
\begin{eqnarray}
\mathcal{L}_{0}\phi(y)=f^{2}(y)-\overline{\sigma}^{2},
\end{eqnarray}
then we have a Poisson equation for which $\phi(y)$ is a solution
to it. We can therefore express equation (\ref{C2}) as
\begin{eqnarray}\label{C2,2}
C_{2}=-\dfrac{1}{2}(\phi(y)+k(s,x))x^{2}\dfrac{\partial^{2}C_{0}}{\partial
x^{2}},
\end{eqnarray}
where k(s,x) is some constant with respect to y, that may depend
on (s,x). %p94,5.30.
%%%%%%%%%%%%%%%%%%%%%%%%%%%%%%%%%%%%%%%%%%%%%%%%%%%%%%%%%%%%%%%%%%%%%%%%%%%%%%%%%%%%%%%%%%%%%%%%%%%%%%%%%%%%%%%%%%%%
\begin{comment}
%********************************DON'T UNDERSTAND, (solves 2nd order linear diff eqn)**************************************************************
% don't think need explicit \psi' expression, Kristine Andersson not cover

We solve the Poisson equation in $\psi(Y)$
$\mathcal{L}_{0}\psi=f^{2}(Y)-\overline{\sigma}^{2}$ to give a
solution to $\psi$ as \cite{fouque2000dfm}:% p94 eqn 5.31
$$\psi'(z)=\dfrac{1}{v^{2}\Phi(z)}\int_{-\infty}^{y} dz ???$$

\end{comment}
%%%%%%%%%%%%%%%%%%%%%%%%%%%%%%%%%%%%%%%%%%%%%%%%%%%%%%%%%%%%%%%%%%%%%%%%%%%%%%%%%%%%%%%%%%%%%%%%%%%%%%%%%%%%%%%%%%%%

Now recall that we have:
\begin{eqnarray}
\mathcal{L}_{0}C_{3}+\mathcal{L}_{1}C_{2}+\mathcal{L}_{2}C_{1} &=&
0\\ \label{C3 Poisson eqn}
 \mbox{so that } \mathcal{L}_{0}C_{3} &=&
-(\mathcal{L}_{1}C_{2}+\mathcal{L}_{2}C_{1}).
\end{eqnarray}
Equation (\ref{C3 Poisson eqn}) is another Poisson equation but in $C_{3}$, %with respect to $\mathcal{L}_{0}$
consequently we can apply the centering condition:
\begin{eqnarray}
<\mathcal{L}_{1}C_{2}+\mathcal{L}_{2}C_{1}> &=& 0\\ \label{L2C1
eqn}
  <\mathcal{L}_{2}C_{1}>&=& -<\mathcal{L}_{1}C_{2}>,\mbox{ since $<>$ is an integral.}
\end{eqnarray}
Now %p94
\begin{enumerate}
\item $C_{1}$ is not a function of y and it was already mentioned
$<\mathcal{L}_{2}>=\mathcal{L}_{BS}(\overline{\sigma})$ so we have
\begin{eqnarray}
<\mathcal{L}_{2}C_{1}> &=&  <\mathcal{L}_{2}>C_{1}\\
&=& \mathcal{L}_{BS}(\overline{\sigma})C_{1}.
\end{eqnarray}
\item substituting our expression for $C_{2}$ from equation (\ref{C2,2}), we have:
\begin{eqnarray}
-<\mathcal{L}_{1}C_{2}>=
\dfrac{1}{2}<\mathcal{L}_{1}\phi(y)>x^{2}\dfrac{\partial^{2}C_{0}}{\partial
x^{2}}.
\end{eqnarray}
Note that the term k(s,x) disappears since $\mathcal{L}_{1}$ takes
derivatives on y terms only.
\end{enumerate}
Using equation (\ref{L2C1 eqn}) we can equate
$-<\mathcal{L}_{1}C_{2}>$ and $<\mathcal{L}_{2}C_{1}>$:
\begin{eqnarray}
\mathcal{L}_{BS}(\overline{\sigma})C_{1}
%&=&   \dfrac{1}{2}<(\mathcal{L}_{1}(\phi(y)+c(t,x))>x^{2}\dfrac{\partial^{2}C_{0}}{\partial x^{2}}\\
    &=&
   \dfrac{1}{2}<\mathcal{L}_{1}\phi(y)>x^{2}\dfrac{\partial^{2}C_{0}}{\partial
   x^{2}}. \label{LBS}
\end{eqnarray}

Now recalling the  $\mathcal{L}_{1}$ operator definition (equation
(\ref{L1 operator definition}) we can determine
$<\mathcal{L}_{1}\phi(y)>$:
\begin{align} %see Kristine Andersson
<\mathcal{L}_{1}\phi(y)> &=& <\left(\sqrt{2}\rho
vf(y)x\dfrac{\partial^{2}}{\partial x \partial
y}-\sqrt{2}v\Lambda(y)\dfrac{\partial}{\partial y} \right)\phi(y)>\\
&=& <\sqrt{2}\rho vf(y)\phi'(y)x\dfrac{\partial}{\partial x}-\sqrt{2}v\Lambda(y)\phi'(y)>\\
&=& \sqrt{2}\rho v<f(y)\phi'(y)>x\dfrac{\partial}{\partial
x}-\sqrt{2}v<\Lambda(y)\phi'(y)>.\label{L1}
\end{align}
We insert equation (\ref{L1}) into equation (\ref{LBS}) giving:
\begin{eqnarray*}
\mathcal{L}_{BS}(\overline{\sigma})C_{1} &=&
   \dfrac{1}{2}<(\mathcal{L}_{1}(\phi(y))>x^{2}\dfrac{\partial^{2}C_{0}}{\partial x^{2}}\\
 &=&
\left( \dfrac{1}{2}\sqrt{2}\rho
v<f(y)\phi'(y)>x^{3}\dfrac{\partial^{3}C_{0}}{\partial x^{3}}
   + \sqrt{2}\rho v<f(y)\phi'(y)> x^{2}\dfrac{\partial^{2}C_{0}}{\partial x^{2}} \right)\\
    &-& \frac{\sqrt{2}}{2}v<\Lambda(y)\phi'(y)>x^{2}\dfrac{\partial^{2}C_{0}}{\partial x^{2}}\\
%NOTE:
%we apply product rule to determine (x\dfrac{\partial}{\partial x})(x^{2}\dfrac{\partial^{2}C_{0}}{\partial x^{2}})
%=x\dfrac{\partial^{3}C_{0}}{\partial x^{3}}+2x^{2}\dfrac{\partial^{2}C_{0}}{\partial x^{2}}
    &=&
    \dfrac{1}{2}\sqrt{2}\rho v<f(y)\phi'(y)>x^{3}\dfrac{\partial^{3}C_{0}}{\partial x^{3}}\\
   &+& \left( \sqrt{2}\rho v<f(y)\phi'(y)>-\frac{\sqrt{2}}{2}v<\Lambda(y)\phi'(y)> \right)x^{2}\dfrac{\partial^{2}C_{0}}{\partial
   x^{2}}.
\end{eqnarray*}
Recalling $\alpha= 1/\epsilon$ we can re-express
$\mathcal{L}_{BS}(\overline{\sigma})(\sqrt{\epsilon} C_{1})$ in a
more convenient form:
\begin{eqnarray}\label{P1 pde}
\mathcal{L}_{BS}(\overline{\sigma})(\sqrt{\epsilon} C_{1})
%&=& (\sqrt{\epsilon})\dfrac{1}{2}(\sqrt{2}\rho v<f(y)\phi'(y)>x\dfrac{\partial}{\partial x}\\
%   &-& \sqrt{2}v<\Lambda(y)\phi'(y)>)x^{2}\dfrac{\partial^{2}C_{0}}{\partial x^{2}}\\
    &=& V_{2}x^{2}\dfrac{\partial^{2}C_{0}}{\partial x^{2}}+V_{3}x^{3}\dfrac{\partial^{3}C_{0}}{\partial
    x^{3}},
\end{eqnarray}
where
\begin{eqnarray}
V_{2} &=& \dfrac{v}{\sqrt{2 \alpha}}(2\rho<f(y)\phi'(y)>-<\Lambda \phi'(y)>),\\
V_{3} &=& \dfrac{\rho v}{\sqrt{2 \alpha}}<f(y)\phi'(y)>,
\end{eqnarray}
with terminal condition $C_{1}(T,x)=0$.
%******************************************************DON'T UNDERSTAND WHY BELOW TRUE, P95****************
 It can be shown the solution to $\sqrt{\epsilon}C_{1}$ gives
\cite{fouque2000dfm}:
%The solution to the right hand side of equation \ref{P1 pde} (the
%partial differential equation) multiplied by -(T-s) to satisfy the
%terminal condition $C_{1}(T,x)=0$ is solved by Fouque
%\cite{fouque2000mrs} %p33
%to give $\sqrt{\epsilon} C_{1}$:%\cite{fouque2000dfm}
\begin{eqnarray}\label{sqrtepsilonC1 eqn}
\sqrt{\epsilon}C_{1}=-(T-s)\left(V_{2}x^{2}\dfrac{\partial^{2}C_{0}}{\partial
x^{2}}+V_{3}x^{3}\dfrac{\partial^{3}C_{0}}{\partial
    x^{3}}\right).
\end{eqnarray}
Therefore:
\begin{eqnarray}
C & = & C_{0}+\sqrt{\epsilon}C_{1}+ \epsilon C_{2}+....\\
& \simeq & C_{0}+\sqrt{\epsilon}C_{1}\\\label{Fouque Final Option
Eqn}
  & \simeq & C_{0}-(T-s)\left(V_{2}x^{2}\dfrac{\partial^{2}C_{0}}{\partial
x^{2}}+V_{3}x^{3}\dfrac{\partial^{3}C_{0}}{\partial
    x^{3}}\right).
\end{eqnarray}
As stated by Fouque \cite{fouque2000dfm} detailed expressions of
$V_{2},V_{3}$ are not important. This %p96
fact is further explained below.
\\\\
\textbf{Option Pricing Using Observed Data}
%\textbf{Fouque's Option Pricing Calibration Method}
\label{Fouque Option Param Est Method}\\
The terms $V_{2},V_{3}$ become explicit functions of the
stochastic volatility if we fully define f in f(Y). However,
Fouque proves that we can price C (specifically equation
(\ref{Fouque Final Option Eqn})) without specifying the stochastic
volatility dynamics by using empirically observed data. To price C
from equation (\ref{Fouque Final Option Eqn}) we need to calculate
$C_{0},V_{2},V_{3},\partial^{2}C_{0}/\partial x^{2}$ and
$\partial^{3}C_{0}/\partial x^{3}$. The variable $C_{0}$ is
calculated using the standard Black-Scholes equation (where
volatility would be $\overline{\sigma}$), variables
$\partial^{2}C_{0}/\partial x^{2}$ and $\partial^{3}C_{0}/\partial
x^{3}$ are known as the option's Gamma and Epsilon respectively
and are calculated by
\begin{eqnarray}\label{Gamma eqn}
\dfrac{\partial^{2}C_{0}}{\partial
x^{2}}=\dfrac{e^{\frac{-d_{1}^{2}}{2}}}{x\overline{\sigma}\sqrt{2\pi(T-s)}}
\end{eqnarray}
and
\begin{eqnarray}\label{Epsilon eqn}
\dfrac{\partial^{3}C_{0}}{\partial
x^{3}}=\dfrac{-e^{\frac{-d_{1}^{2}}{2}}}{x^{2}\overline{\sigma}\sqrt{2\pi(T-s)}}\left(1+
\dfrac{d_{1}}{\overline{\sigma}\sqrt{T-s}} \right),
\end{eqnarray}
where $d_{1}$ is taken from the Black-Scholes equation.

The variables $V_{2},V_{3}$ are found by fitting an affine
function to a logarithmic plot of empirical option prices.
%$$C_{BS}(t,x,K,T,I)=C_{observed}(K,T)$$
To prove this, we first recall that implied volatility
$\mathcal{I}$ is given by
$$C_{BS}(x,s,T,r,\mathcal{I},K)=C^{obs},$$
where $C_{BS}$ is the European call option price under
Black-Scholes option pricing. Now $C^{obs}$ is modelled by our
perturbation expansion, that is:
\begin{eqnarray}
C_{BS}(x,s,T,r,\mathcal{I},K) &=& C^{obs}\\
                              &=& C_{0}+\sqrt{\epsilon}C_{1}+... .
\end{eqnarray}

We can expand $\mathcal{I}$ as:
\begin{eqnarray}\label{I expansion}
\mathcal{I} &=& \mathcal{I}_{0}+\sqrt{\epsilon}\mathcal{I}_{1}+...
\end{eqnarray}
We can also use a Taylor series expansion to expand function
$C_{BS}$. A function f(x) expanded around x=k by Taylor series
expansion gives \cite{kreyszig:aem}:
\begin{eqnarray}
f(x)=f(k)+f'(k)(x-k)+\dfrac{f''(k)(x-k)^{2}}{2!}+...+\dfrac{f^{n}(k)(x-k)^{n}}{n!}.
\end{eqnarray}
Therefore a Taylor series expansion of
$C_{BS}(x,s,T,r,\mathcal{I},K)$ around
$\mathcal{I}=\mathcal{I}_{0}$ gives
\begin{eqnarray}
C_{BS}(x,s,T,r,\mathcal{I},K) &=&
C_{BS}(x,s,T,r,\mathcal{I}_{0},K)\\
&+&(\mathcal{I}-\mathcal{I}_{0})\dfrac{\partial
C_{BS}(x,s,T,r,\mathcal{I}_{0},K)}{\partial \sigma}+...\\
\label{CBS expansion eqn}
 &=&
C_{BS}(x,s,T,r,\mathcal{I}_{0},K)+\sqrt{\epsilon}\mathcal{I}_{1}\dfrac{\partial
C_{BS}(x,s,T,r,\mathcal{I}_{0},K)}{\partial \sigma}\\
&+&...  .
\end{eqnarray}
The last line was possible since
$\sqrt{\epsilon}\mathcal{I}_{1}=\mathcal{I}-\mathcal{I}_{0}$ from
equation (\ref{I expansion}).

Equating equation (\ref{CBS expansion eqn}) with expansion
$C=C_{0}+\sqrt{\epsilon}C_{1}+..$ we have:
\begin{eqnarray}
C_{BS}(x,s,T,r,\mathcal{I}_{0},K)+\sqrt{\epsilon}\mathcal{I}_{1}\dfrac{\partial
C_{BS}(x,s,T,r,\mathcal{I}_{0},K)}{\partial \sigma}+...
=C_{0}+\sqrt{\epsilon}C_{1}+... .
\end{eqnarray}
By standard perturbation analysis we equate terms of the same
order; matching terms of order $\mathcal{O}(1)$ we have:
\begin{eqnarray*}
C_{BS}(x,s,T,r,\mathcal{I}_{0},K) &=& C_{0} \Rightarrow
\mathcal{I}_{0} = \overline{\sigma}.
\end{eqnarray*}

Equating terms of order $\mathcal{O}(\sqrt{\epsilon})$ gives
\begin{eqnarray}
\sqrt{\epsilon}\mathcal{I}_{1} &=&
\sqrt{\epsilon}C_{1}\left[\dfrac{\partial
C_{BS}(x,s,T,r,\overline{\sigma},K)}{\partial \sigma}
\right]^{-1}.
\end{eqnarray}
If we now insert this expression for
$\sqrt{\epsilon}\mathcal{I}_{1}$ into our expansion for
$\mathcal{I}$ we have:
\begin{eqnarray}
\mathcal{I} &=&
\overline{\sigma}+\sqrt{\epsilon}\mathcal{I}_{1}+\mathcal{O}(\epsilon)\\
\label{IV eqn}
 &=&
 \overline{\sigma} + \sqrt{\epsilon}C_{1}\left[\dfrac{\partial
C_{BS}(x,s,T,r,,\overline{\sigma},K)}{\partial \sigma}\right]^{-1}
+\mathcal{O}(\epsilon).
\end{eqnarray}
We can also make the substitution for $\frac{\partial
C_{BS}}{\partial \sigma}$ in equation (\ref{IV eqn}) using the
option's Vega:
\begin{eqnarray}
\dfrac{\partial C_{BS}}{\partial \sigma} =
\dfrac{xe^{\frac{-d_{1}^{2}}{2}}\sqrt{T-s}}{\sqrt{2\pi}}.
\end{eqnarray}
Now substituting equations (\ref{Gamma eqn}) and (\ref{Epsilon
eqn}) into equation (\ref{sqrtepsilonC1 eqn}) and after some
rearranging, it can be shown we have:
\begin{eqnarray}
\mathcal{I} &=&
\textit{aL}+\textit{b}+\mathcal{O}(\epsilon),\label{I=aL+b eqn}
\end{eqnarray}
where
\begin{itemize}
\item \textit{L}=$\dfrac{log(K/x)}{(T-s)}$;

\item $\textit{a}=-\dfrac{V_{3}}{\overline{\sigma}^{3}} \Rightarrow
V_{3}=-a\overline{\sigma}^{3}$;

\item  $\textit{b}=\overline{\sigma}+
\dfrac{V_{3}}{\overline{\sigma}^{3}}\left(r+\dfrac{3}{2}\overline{\sigma}^{2}\right)-\dfrac{V_{2}}{\overline{\sigma}}
\Rightarrow
V_{2}=\overline{\sigma}((\overline{\sigma}-b)-a(r+\frac{3}{2}\overline{\sigma}^{2}))$.
\end{itemize}
%For calibration purposes Fouque uses the following:
%$$V_{2}=\overline{\sigma}((\overline{\sigma}-b)-a(r+\frac{3}{2}\overline{\sigma}^{two}))$$
%$$V_{3}=-a\overline{\sigma}^{3}$$
We can therefore obtain \textit{a,b} from a simple linear plot of
equation (\ref{I=aL+b eqn}) (given we know r and
$\overline{\sigma}$). Plotting $\mathcal{I}$ on the y-axis and
(log(K/x)/T) on the x-axis, the gradient gives \textit{a} and the
y-intercept is \textit{b} (in fact we can
think of \textit{b} as the at the money $\mathcal{I}$). %(note:
%Fouque fits line by minimum least squares difference \cite{fouque2000mrs}p25).

We know that any function of Y f(Y) (that is any mean reverting
stochastic volatility process) tends to a constant
$\overline{\sigma}^{2}=<f^{2}(Y)>$ and  $\overline{\sigma}$ can be
measured from empirical price data without specifying the
volatility's dynamics. Methods of measuring $\overline{\sigma}$
are not required but the reader is referred to
\cite{fouque2000dfm} for more information.

We can now determine stochastic volatility option prices under
mean reversion without specifying the volatility's dynamics. The
variable $\overline{\sigma}$ is obtained from price data,
variables required to determine $V_{2},V_{3}$ are observable (from
the linear plot of equation (\ref{I=aL+b eqn})). Therefore
Fouque's option pricing method is model independent;  all that is
required is that the stochastic volatility process is a function
of Y (mean reverting).
%and such that they admit a solution in the Poisson
%equations.%\cite{fouque2000dfm}p96s

We note that the option pricing accuracy is of the order of
$\mathcal{O}(\epsilon)$: %\cite{fouque2000dfm} p102
\begin{eqnarray}
|C-(C_{0}+\sqrt{\epsilon}C_{1})|\approx \mathcal{O}(\epsilon).
\end{eqnarray}
Since $\alpha=\frac{1}{\epsilon}$ and we have assumed fast mean
reversion we can assume the pricing error is negligible. For
higher order approximations the option pricing method is no longer
model independent and we would require specific knowledge of the
stochastic volatility dynamics.

\subsection{Regime Switching Perturbation Based Option Pricing}\label{Regime Switching and Pertubation Based Option
Pricing}
%argument organisation: regime (with rs-mrsv) has dist (and
%constant vol assoc) - relate to constant vol in fouque. prove
Regime switching models generally specify the regime's
distribution but not the intra-regime dynamics. Consequently if we
can price intra-regime options without explicit knowledge of the
intra-regime dynamics (just assuming mean reverting stochastic
volatility) it would be applicable to general RSMR models.

Now Fouque's option pricing method is model independent as it does
not require any specific definition of the mean reverting
stochastic volatility process to price options. All the pricing
variables are either exogenously determined variables (such as the
strike and expiry) or are empirically observable (such as the
stock price and interest rates), except $\bar{\sigma}$, which is
calculated from stock price data. Therefore if we can relate
$\bar{\sigma}$ to a typical regime switching model's specification
we can obtain intra-regime option prices for general regime
switching models, assuming a mean reverting stochastic volatility
intra-regime process. We will now demonstrate this.
%%%%%%%%%%%%%%%%%%%%%%%%%%%%%%%%%%%%%%%%%%%%%%%%%%%%%%%%%%%%%%%%%%%%%%%%%%%%%%%%%%%%%%%%%%%%%%%%%%%%%%%%%%%%%%%
\begin{comment}
for any regime switching model we specify for each regime R a
constant volatility for the duration of the regime. Our aim is to
determine the relation between the mean reverting stochastic
volatility process and this constant volatility.

In assigning a return distribution we are effectively assigning a
constant volatility equivalent to each regime,
$\overline{\sigma}_{R}$, which does not describe the ''true"
volatility dynamics. We can interpret $\overline{\sigma}_{R}$ as
if we were writing the stochastic differential equation as
$$dX/X=\mu_{R}dt+\overline{\sigma}_{R}dW, R \in \{1,2\}$$
Generally in regime switching models the intra-regime dynamics are
not related to $\overline{\sigma}_{R}$. We will now show how this
can be achieved.
\end{comment}
%%%%%%%%%%%%%%%%%%%%%%%%%%%%%%%%%%%%%%%%%%%%%%%%%%%%%%%%%%%%%%%%%%%%%%%%%%%%%%%%%%%%%%%%%%%%%%%%%%%%%%%%%%%%%%%

If the intra-regime dynamics had simply been time dependent
volatility $\sigma(t)$, an analytic relation exists between
$\sigma(t)$ and the return distribution. If we denote the constant
volatility equivalent of $\sigma(t)$ by $\sigma_{c}$
\begin{eqnarray}\label{constant vol eqn Fouque}
\sigma_{c}=\sqrt{\dfrac{1}{T-t}\int^{T}_{t}\sigma^2(\tau)d\tau},
\end{eqnarray}
then the distribution would be given by:
\begin{eqnarray}\label{time dep dist}
log(X(T)/X(t))\sim \mathcal{N}\left(\left(\mu-
\frac{1}{2}\sigma_{c}^{2} \right)(T-t),\sigma_{c}^{2}(T-t)\right).
\end{eqnarray}
Our approach will be to demonstrate a similar relation for mean
reverting stochastic volatility.
%%%%%%%%%%%%%%%%%%%%%%%%%%%%%%%%%%%%%%%%%%%%%%%%%%%%%%%%%%%%%%%%%%%%%%%%%%%%%%%%%%%%%%%%%%%%%%%%%%%%%%%%%%%%%%%%
\begin{comment}
to $\overline{\sigma}_{R}$ using the time dependent volatility
model discussed in section \ref{Time Dependent Volatility Model}:
\begin{eqnarray}
{\overline{\sigma}}^{2}_{R}=\dfrac{1}{T}\int^{T}_{0}\sigma^2(\tau)d\tau,
\forall R.
\end{eqnarray}
Here T denotes the duration of 1 regime before switching occurs (1
year). For intra-regime dynamics consisting of local volatility
$\sigma(X,t)$ there is no known formula that relates it to the
constant volatility equivalent $\overline{\sigma}_{R}$.
%In stochastic volatility where the Wieners are uncorrelated ($\rho=0$) we can apply the the Hull-White formula to relate the
%regime's constant volatility. -NOT SURE???

Fouque relates a constant volatility equivalent to the stochastic
volatility process provided it is highly mean reverting. For
regime switching models with mean reverting stochastic volatility
intra-regime processes (e.g. RSexpOU) the constant volatility
therefore represents the regime's volatility
$\overline{\sigma}_{R}$. We will now prove this. Note that
Fouque's method is not strictly concerned with regime switching,
therefore we will not discuss it with reference to regime
switching unless specifically mentioned.
\end{comment}
%%%%%%%%%%%%%%%%%%%%%%%%%%%%%%%%%%%%%%%%%%%%%%%%%%%%%%%%%%%%%%%%%%%%%%%%%%%%%%%%%%%%%%%%%%%%%%%%%%%%%%%%%%%%%%%%

In section \ref{Perturbation Formulation of Stochastic Volatility
Option Pricing} we defined a mean reverting stochastic volatility
process where $\alpha$ is the rate of mean reversion and is
empirically observed to be very high ($\alpha \rightarrow\infty$).
%\cite{fouque2000dfm}p66
%Fouque Method requires high rate of mean reversion $\alpha$ for a
%chosen time scale. What does this mean? for stock prices $\alpha$
%is low when chosen in minutes time scale, for months $\alpha$ is
%high (\cite{fouque2000mrs}abstract,p8). Hence over months we
%expect prices to revert around the mean frequently but not in
%minutes time scale. Naturally, for a given time scale, if we
%increase $\alpha$ we expect reversion to increase -cf p.59 figure
%3.2 \cite{fouque2000dfm}. Thus from empirical data we need to
%validate $\alpha$ is large.
A mean reverting stochastic volatility process $\sigma=f(Y)$ has
the important property of ergodicity and it is this property which
enables us to link a regime's distribution to its intra-regime
dynamics. For a more detailed review of ergodicity the reader is
referred to Kannan \cite{kannan1979isp}.

To explain ergodicity let g(X) be a process, where g is a function
of an ergodic process X(t). The expectation for g(X) is
\begin{eqnarray}
E[g(X)] =\int_{-\infty}^{\infty}g(X)p(X)dX.
\end{eqnarray}
Note that the expectation is generally a function of time. For an
ergodic process X(t) or a function of an ergodic process g(X)
there exists an invariant distribution, which is the long term or
equilibrium distribution exhibited by g(X) as time tends to a
large limit. If we denote $\Phi(\cdot)$ as the probability measure
of the invariant distribution for X(t) then the ensemble average
(also known as the statistical average) is the expectation
under $\Phi(\cdot)$ probability  measure:%Fouque \cite{fouque2000dfm,p69}:
\begin{eqnarray}
E^{\Phi}[g(X)]=\int_{-\infty}^{\infty}g(X)\Phi(X)dX.
% NOT SURE SO LEAVE:
%\mbox{and}\\
%E[g(Y_{t})|Y_{0}]\rightarrow <g(Y_{t})>, t\rightarrow \infty%p63
\end{eqnarray}
Now the time average of $\overline{g(X)}$ is defined as:
\begin{eqnarray}
\overline{g(X)}=\dfrac{1}{t}\int^{t}_{0}{g(\tau)}d\tau.
\end{eqnarray}
Note that a time average is generally a random variable. We define
the time average $<g(X)>$ as:
\begin{eqnarray}
<g(X)>=\lim_{t\rightarrow \infty} \dfrac{1}{t} \int_{0}^{t}g(X)ds.
\end{eqnarray}
We say X(t) is an ergodic process if
\begin{eqnarray}\label{time equal ens avg}
E^{\Phi}[g(X)] =<g(X)>,
\end{eqnarray}
that is X(t) is an ergodic process if g(X) has the property that
the time average as $t\rightarrow \infty$ or $<g(X)>$ equals its
ensemble average under its probability measure $\Phi(X)$. Note
that in equation (\ref{time equal ens avg}) both $<g(X)>$ and
$E^{\Phi}[g(X)]$ must always be a constant for the equation to hold. %(also true by the property of ergodicity discussed in \ref{RSexpOU Baum-Welch
%Calibration by Gaussian Mixture HMM}).

Now for mean reverting stochastic volatility, $\sigma=f(Y)$ is
ergodic therefore its time average approaches $<f(Y)>$ (and this
is a constant):
\begin{eqnarray}
\lim_{t\rightarrow \infty} \dfrac{1}{t} \int_{0}^{t}f(Y)ds=<f(Y)>.
\end{eqnarray}
This relation is true regardless of the value of $\alpha$. If %p66-7 Fouque bk
 t is large but finite and $\alpha\rightarrow \infty$ then we
have: %Fouque bk p67
\begin{eqnarray}
\dfrac{1}{t} \int_{0}^{t}f(Y)ds \approx <f(Y)>.
\end{eqnarray}
Therefore the time average of a highly mean reverting stochastic
volatility process approaches the constant $<f(Y)>$.
%%%%%%%%%%%%%%%%%%%%%%%%%%%%%%%%%%%%%%%%%%%%%%%%%%%%%%%%%%%%%%%%%%%%%%%%%%%%%%%%%%%%%%%%%%%%%%%%%%%%%%%%%%%%%%%%
\begin{comment}
Since the SDE changes for each regime so will $\overline{\sigma}$,
so it is more correct to write $\overline{\sigma}_{R}$. We could
approximate the SDE as
\begin{eqnarray}
dX/X\approx \mu_{R}dt+\overline{\sigma}_{R}dW, R \in \{1,2\}
\end{eqnarray}
In other words under fast mean reversion it is \textit{as though}
we have constant volatility, as in the Black-Scholes equation
derivation.

The constant volatility (equivalent) for each regime
$\overline{\sigma}_{R}$ can be determined by calculating equation
\ref{time equal ens avg} for each regime. If the intra-regime
dynamics are explicitly known then this can be done analytically.
For example....????

If the distribution is known only then for each regime since
($\overline{\sigma}_{R}$) is a constant, it is the standard
deviation of returns.

\end{comment}
%%%%%%%%%%%%%%%%%%%%%%%%%%%%%%%%%%%%%%%%%%%%%%%%%%%%%%%%%%%%%%%%%%%%%%%%%%%%%%%%%%%%%%%%%%%%%%%%%%%%%%%%%%%%%%%%

The ergodic relation of equation (\ref{time equal ens avg}) is
also
true for $g^{2}(X)$ \cite{fouque2000dfm}: %p67
\begin{eqnarray}
<g^{2}(X)>=\lim_{t\rightarrow \infty} \dfrac{1}{t}
\int_{0}^{t}g^{2}(X)ds.
\end{eqnarray}
Therefore the time average of $\sigma^{2}=f^{2}(Y)$ approaches
$<g^{2}(Y)>$ (a constant):
\begin{eqnarray}
\lim_{t\rightarrow \infty} \dfrac{1}{t} \int_{0}^{t}f^{2}(Y)ds &=&
<f^{2}(Y)>.
\end{eqnarray}
We also defined in section \ref{Perturbation Solution to Option
Pricing section}:
\begin{eqnarray}
 (\overline{\sigma})^{2} &=& <f^{2}(Y)>,\\
 &=& \lim_{t\rightarrow \infty} \dfrac{1}{t}
 \int_{0}^{t}f^{2}(Y)ds.
\end{eqnarray}
If t is finite but $\alpha\rightarrow \infty$ then we have:
\begin{eqnarray}
\dfrac{1}{t} \int_{0}^{t}f^{2}(Y)ds & \approx & <f^{2}(Y)>,\\
& \approx &  \overline{\sigma}^{2}.
\end{eqnarray}
We also know that it is possible to convert time dependent
volatility to its constant volatility equivalent $\sigma_{c}$ from
section \ref{Time Dependent Volatility Model}, that is equation
(\ref{constant vol eqn Fouque}). The constant volatility
equivalent $\sigma_{c}$ for f(Y) is therefore approximated  by:
\begin{eqnarray}
\sigma_{c} &=& \sqrt{\frac{1}{t}\int_{0}^{t}f^{2}(Y(s))ds},\\
\sigma_{c} & \approx & \sqrt{<f^{2}(Y(s))>},\\
& \approx & \overline{\sigma}.
\end{eqnarray}
Therefore applying equation (\ref{time dep dist}) with $\sigma_{c}
\approx \overline{\sigma}$, we have a way relating any general
mean reverting stochastic volatility process to the return
distribution: $\overline{\sigma}$ is the regime's standard
deviation. The distribution of each regime is approximated by
equation (\ref{time dep dist}) with $\sigma_{c} \approx
\overline{\sigma}$ and as $t \rightarrow \infty$ the approximation
is precise. The variable $\overline{\sigma}$ is specific to each
regime since with each regime the volatility parameter values may
change (and therefore the return distribution); we therefore write
$\overline{\sigma}_{i}$ where $i \in \{1,..,R\}$.

Now as we are pricing intra-regime options we do not price options
with expiries beyond the date of the next possible regime change.
Consequently, for intra-regime options the regime does not change
during the option's life and so we do not need to take into
account risk arising from switching regimes. Therefore regimes do
not introduce any additional incompleteness into intra-regime
option pricing. Furthermore, $\overline{\sigma}$ takes on only one
value during the life of an option and so we can obtain
intra-regime option prices using Fouque's option pricing method.

Now as we have established Fouque's option pricing method is model
independent, $\overline{\sigma}_{i}$ (where $i \in \{1,...,R\}$)
is specified for general regime switching models and assuming we
have RSMR, we can therefore price intra-regime options for any
general RSMR model. Since we have
\begin{eqnarray}
C &\simeq& C_{0}+\sqrt{\epsilon}C_{1}+ \epsilon C_{2}+....,
\end{eqnarray}
our intra-regime option pricing equation for RSMR is therefore:
\begin{eqnarray}\label{RSMR eqn1}
C_{i} &=& C_{0i}+\sqrt{\epsilon}C_{1i}+ \epsilon C_{2i}+....\\
\label{RSMR eqn2}
 & \simeq & C_{0i}+\sqrt{\epsilon}C_{1i}, i \in \{1,...,R\},
\end{eqnarray}
where $\overline{\sigma}_{i}$ is specified for each regime.

If we wished to price options at any point in time without
knowledge of the current state and  beyond the next state
transition then we can apply Boyle's and Draviam's regime
switching option pricing equation \cite{boyle2007peo}, which is a
transition probability weighted sum of option prices. However,
long dated options (options with expiries longer than the duration
of one regime (one year)) tend to be rarely traded, consequently
their prices are significantly distorted by illiquidity effects
(to be discussed in section \ref{Fouque Calibration Procedure}).
In fact many traded and liquid options expire at times
considerably less than one year, hence the limit on option expiry
is not restrictive.
%%%%%%%%%%%%%%%%%%%%%%%%%%%%%%%%%%%%%%%%%%%%%%%%%%%%%%%%%%%%%%%%%%%%%%%%%%%%%%%%%%%%%%%%%%%%%%%%%%%%%%%%%%%%%%%%%
\begin{comment}
However the purpose of our model (and numerical experiments) is to
demonstrate that a regime specific $\overline{\sigma}_{i}$
improves option pricing and we know that only one regime is
occupied any point in time, so it is better to compare results for
the (most likely) correct regime with respect to empirical data
without factoring in uncertainty of the current regime.
Furthermore, as stated by Bollen
\cite{bollen1998vor} %p11
regimes reflect economic or long term fundamentals so it is
possible to detect the correct regime from data (although
mathematically regimes are hidden Markov states), therefore the
option prices are likely to reflect the correct current regime
rather than mispricing under another regime.
\end{comment}
%%%%%%%%%%%%%%%%%%%%%%%%%%%%%%%%%%%%%%%%%%%%%%%%%%%%%%%%%%%%%%%%%%%%%%%%%%%%%%%%%%%%%%%%%%%%%%%%%%%%%%%%%%%%%%%%%
\\\\
\textbf{Option Pricing Advantages}\\
 We now discuss
the advantages of our  option pricing method. Firstly, current
regime switching option pricing methods tend to neglect
intra-regime dynamics but also are model specific. Intra-regime
option pricing using Fouque's perturbation approach is applicable
to general regime switching models without requiring explicit
knowledge of intra-regime dynamics. All that is required is that
we know $\bar{\sigma}_{i}$ and we assume  mean reverting
stochastic volatility intra-regime dynamics (since Fouque's option
pricing method is model independent).

A possible method of intra-regime option pricing would be to apply
the appropriate option pricing formula to the intra-regime
dynamics. However, this requires explicit knowledge of the
intra-regime process and is not strictly regime based option
pricing since such a method is not purely a function of the regime
switching model. There is also the significant disadvantage that
Wiener process driven stochastic volatility models with correlated
Wiener processes have no analytic solutions, excluding the Heston
model. Furthermore, the Heston model assumes a specific stochastic
volatility process, risk neutral measure and there is no stated
relation between the volatility of a regime switching model and
the Heston model's volatility process.

Secondly, our  method  assumes mean reverting stochastic
volatility intra-regime dynamics. The assumption of mean reverting
stochastic volatility intra-regime dynamics is less restrictive
and more realistic compared to local and time dependent volatility
intra-regime dynamics. Hence such an intra-regime option pricing
approach should provide better option pricing compared to other
intra-regime or general regime switching pricing methods.

Thirdly, in contrast to other stochastic volatility option pricing
methods or regime switching option pricing methods with no
intra-regime dynamics, Fouque does not choose some particular
function for specifying a unique risk neutral
measure. Instead  % \cite{fouque2000mrs}p11.
 Fouque uses the ``market's view" of the market price of
volatility risk by plotting equation (\ref{I=aL+b eqn}) to extract
a unique risk neutral measure from empirical option data.
Consequently our intra-regime option pricing can be calibrated to
current option data and is not restricted by any specific risk
neutral measure.
%Once the equation is in singular perturbation form we can solve it
%using many methods but the method employed in \cite{fouque2000dfm}
%is the \textit{method of matched asymptotic series expansions}.

Fourthly, an important advantage of Fouque's option pricing method
is that calibration is far more parsimonious compared to other
stochastic volatility option pricing methods. All that is required
is $\bar{\sigma}$ and a linear fit of equation (\ref{I=aL+b eqn});
the calibration is just as parsimonious under regime based Fouque
option pricing where we use $\bar{\sigma}_{i}$ instead. Typically
Wiener process driven stochastic volatility option prices require
calibrating many volatility parameters.

Fifthly, using our knowledge of perturbation analysis we can use
Fouque's method to obtain approximations to intra-regime options
for RSMR. We can express option pricing for RSMR using equations
(\ref{RSMR eqn1}) and (\ref{RSMR eqn2}). From perturbation theory
we know we can therefore approximate $C_{i}$ by a high degree of
accuracy by $C_{0i}$ since $C_{0i}$ is the largest term and the
remaining terms are multiplied by $\epsilon^{n}$, where $n>0$. Now
we know from equation (\ref{LBSC0=0}) that $C_{0}$ is simply the
option price using the Black-Scholes equation therefore $C_{0i}$
is simply the Black-Scholes option price for regime i with
volatility $\sigma=\bar{\sigma}_{i}$. This approximation has the
advantage that we do not require option data for calibration or
the market price of volatility risk.
%%%%%%%%%%%%%%%%%%%%%%%%%%%%%%%%%%%%%%%%%%%%%%%%%%%%%%%%%%%%%%%%%%%%%%%%%%%%%%%%%%%%%%%%%%%%%%%%%%%%%%%%%%%%%%%%%
\begin{comment}-LEAVE BECAUSE DON'T UNDERSTAND WHAT HAPPENS TO
SMILE EFFECTS AS T->INFINITY.

Intuitively we expect option prices to be close to Black-Scholes
option prices with  ${\sigma}=\bar{\sigma}$. From ergodic theory
we know $\overline{\sigma}$ approaches a constant:
\begin{eqnarray}
 \overline{\sigma}^{2} &=& <f^{2}(Y)>,\\
 &=& \lim_{t\rightarrow \infty} \dfrac{1}{t}
 \int_{0}^{t}f^{2}(Y)ds.
\end{eqnarray}
As mentioned previously, we can calculate the constant volatility
equivalent $\sigma_{c}$ from
\begin{eqnarray}
\sigma_{c}^{2} &=& \frac{1}{t}\int_{0}^{t}f^{2}(Y(s))ds,\\
\sigma_{c} &=& \sqrt{\frac{1}{t}\int_{0}^{t}f^{2}(Y(s))ds},\\
\sigma_{c} & = & \sqrt{<f^{2}(Y(s))>},\\
& = & \overline{\sigma}, t\rightarrow \infty.
\end{eqnarray}
Therefore as time increases the constant volatility equivalent
approaches  constant $\bar{\sigma}$ and so $\bar{\sigma}$ becomes
a more accurate approximation of f(Y). Therefore the Black-Scholes
equation with $\sigma=\bar{\sigma}$ should become a better
approximation of option prices under f(Y) as time increases.

%The zero order term $C_{0}$ gives the Black-Scholes option price
%for $\sigma=\bar{\sigma}$ and therefore $C_{0}$ cannot account for
%volatility smiles -the smile is accounted for by higher order
%terms in the perturbation series ($\sqrt{\epsilon}C_{1}$ and
%higher). However as

\end{comment}
%%%%%%%%%%%%%%%%%%%%%%%%%%%%%%%%%%%%%%%%%%%%%%%%%%%%%%%%%%%%%%%%%%%%%%%%%%%%%%%%%%%%%%%%%%%%%%%%%%%%%%%%%%%%%%%%%

The approximation $C_{i} \simeq  C_{0i}$ is advantageous since we
can apply various regime switching option pricing formulae that
typically assume volatility is constant in each regime, for
instance Mamon and Rodrigo \cite{mamon2005ese}. Additionally,
$C_{0i}$ is calculated by the Black-Scholes equation, which is
advantageous because much literature exists on the Black-Scholes
equation and so can be directly applied to RSMR models. In fact
Black-Scholes option pricing is used as a building block for more
complex applications, for example Lieson \cite{leisen1999vbo}
values barrier options (options where the possibility to exercise
depends on the underlying crossing a barrier) with jump risk using
a Black-Scholes framework.

%\textbf{Fouque Pricing Benefits}\\
%add all RS benefits\\
Finally, the addition of regime switching $\bar{\sigma}_{i}$ also
improves Fouque's option pricing method:
\begin{itemize}
  \item Firstly, Fouque calculates $\bar{\sigma}$ from past historic data
assuming no regime switching occurs (that is there is only one
$\bar{\sigma}$ value). This is because Fouque assumes stochastic
volatility parameters do not change with time, which is not
realistic over the long term. In fact it is worth noting that a
current area of research for Fouque option pricing is addressing
nonstationary stochastic volatility (see \cite{fouque2000mrs}).

  \item Secondly,  it can be empirically shown that the expOU parameter values
change with each regime; this is effectively the same as
$\bar{\sigma}$ changing with each regime i. This suggests we can
improve Fouque's option pricing over the long term by introducing
regime switching to model $\bar{\sigma}$.

  \item Thirdly, $\bar{\sigma}$ is fundamental to pricing C;  in fact the
most significant term in the expansion of C, $C_{0}$, is the
Black-Scholes option price with volatility $\sigma=\bar{\sigma}$,
therefore C is sensitive to changes in $\bar{\sigma}$.
Specifically, the regime switching captures how $\bar{\sigma}_{i}$
changes as the economy cycles through various economic phases  and
therefore should provide more accurate option prices.

\end{itemize}

%Models long term and non-stationary\\
%The application of regime switching to stochastic volatility
%offers a tractable method of modelling nonstationary stochastic volatility.

\section{Numerical Experiment: Intra-Regime Option
Pricing}\label{Numerical Experiment:SV Intra-Regime option
pricing}
%%%%%%%%%%%%%%%%%%%%%%%%%%%%%%%%%%%%%%%%%%%%%%%%%%%%%%%%%%%%%%%%%%%%%%%%%%%%%%%%%%%%%%%%%%%%%%%%%%%%%%%%%%%%%%%
\begin{comment}
IMPORTANT:\\
ONLY TESTED 4 DATES BUT SAY RESULTS ARE REPRESENTATIVE SO DON'T
NEED TEST ALL.
\end{comment}
%%%%%%%%%%%%%%%%%%%%%%%%%%%%%%%%%%%%%%%%%%%%%%%%%%%%%%%%%%%%%%%%%%%%%%%%%%%%%%%%%%%%%%%%%%%%%%%%%%%%%%%%%%%%%%%

\subsection{Calibration Procedure}\label{Fouque Calibration
Procedure}
 %DON'T NEED DO EXPOU OPTION PRICING BECAUSE SHOULD BE SAME AS CONSTANT VOL
In this section we perform numerical experiments on pricing
options on the S\&P 500 index. The purpose of the experiments is
to firstly demonstrate we can obtain intra-regime option prices
for a general RSMR model, where its regime distributions are
specified but its intra-regime dynamics are not. Secondly, we aim
to compare the option pricing performance against Black-Scholes
pricing and standard Fouque pricing. We demonstrate that by
applying a regime specific $\overline{\sigma}_{i}$ improves
Fouque's option pricing, as opposed to applying one
$\overline{\sigma}$ which is currently proposed by Fouque.

The experimental procedure was executed as follows. We fitted a
linear plot to equation (\ref{I=aL+b eqn}) to extract \textit{a}
and \textit{b} from the set of option prices for the chosen quote
date; for ``standard" Fouque option pricing we used only one
$\overline{\sigma}$ value for any quote period. For the
Black-Scholes option pricing we set volatility
$\sigma=\overline{\sigma}$ as we want to compare the option
pricing performance of the Black-Scholes model for the same
volatility values.

For our ``regime based" Fouque option pricing, to calculate
\textit{a} and \textit{b} in equation (\ref{I=aL+b eqn}) we
applied the appropriate $\overline{\sigma}_{i}$ according to the
regime i of the quote date. We identified the option quote date's
regime using out of sample results from previous experimental
results. To calculate $\overline{\sigma}_{i}$ it was shown in
section \ref{Regime Switching and Pertubation Based Option
Pricing} that $\overline{\sigma}_{i}$ is related to the return
distribution's standard deviation for a given regime; these were
calculated to be in regime 1 as $11.6\%/year$, regime 2 as
$13.3\%/year$. From previous experimental work the non-regime
switching $\overline{\sigma}$ was calculated to be
$\overline{\sigma}=12.77\%/year$.

To compare the option pricing performance under Black-Scholes
pricing, standard Fouque pricing and regime switching based Fouque
pricing, we calculated the average percentage error of their
prices against the empirically observed prices of the S\&P 500
options:
\begin{eqnarray}
\frac{1}{N} \sum_{i=1}^{N}\dfrac{|\mbox{(observed option price) -
(model option price)}|}{\mbox{observed option price}},
\end{eqnarray}
where N is total number of option prices. Note that since regime
switching has been proven to effectively model constant volatility
over long periods, we expect Fouque's method to provide better
results when using $\overline{\sigma}_{i}$ than just
$\overline{\sigma}$.
%
%MENTION?
% PRICE OPTIONS WITH EXPIRY BEFORE NEXT REGIME CHANGE POSSIBLE
%MUST DO BECAUSE V2,V3 INCLUDE \BAR(\SIGMA), WHICH USED TO PRICE OPTIONS FROM OPTION DATA,
%THEREFORE WE CANNOT USE OPTION DATA WITH EXPIRY GREATER THAN NEXT REGIME CHANGE SINCE \BAR(\SIGMA) MAY CHANGE

%\underline{Option Data Issues}\\
%\textit{Equation Accuracy} *p102-4: approx accuracy proof
Numerical experiments performed with options data must be
conducted differently compared to those on equities data. For a
stock we can assign one price to it at a given point in time by
taking the mid-point price of the bid and ask prices (as is
generally practised in industry). A mid-point price is an
acceptable indication of the stock's price since the bid-ask
spread for stocks in general does not significantly vary over
time.

For an option assigning a single price for each point in time is
more problematic than it is for stocks. Firstly, option prices are
a function of a higher number of independent variables than
stocks, so for a given point in time its quoted price will depend
on the strike K and expiry T. Secondly, options are significantly
affected by illiquidity effects due to low trading volume. This
leads to wide variations in bid-ask spreads, which in turn are a
function of K and T (see for instance Pinder \cite{pinder2003eei},
George and Longstaff \cite{george1993bas}). Consequently, there is
no accepted method of assigning a single price to an option from
bid and ask prices.

Following Fouque \cite{fouque2000dfm} to avoid illiquidity effects
we used highly traded options (S\&P 500 options), used near at the
money options with $|K/X-1|\leq 3\%$ and assigned the mid-point
price of bid and ask prices for a
given option price. We selected options data with expiries of at least three weeks % Fouque p110
since options close to expiry tend to be affected by more
speculative and irrational effects. Under such effects Fouque
states that it is doubtful if \textit{any} diffusion model is
useful for short expiries.
%%%%%%%%%%%%%%%%%%%%%%%%%%%%%%%%%%%%%%%%%%%%%%%%%%%%%%%%%%%%%%%%%%%%%%%%%%%%%%%%%%%%%%%%%%%%%%%%%%%%%%%%%%%%%%%%%%
\begin{comment} DON'T UNDERSTAND
Additionally, short expiry (and out of the money) options causes L
in equation \ref{I=aL+b eqn} to become large, where
$$L=\dfrac{log(K/X)}{(T-t)}$$
Therefore
 as small
corrections become
large \cite{fouque2000dfm}.%p105
%\cite{fouque2000mrs}p13. A good approx if
%$(T-t)/|log(K/x)|>>(\alpha)^{-1/2}$ $=>$don't use far out/in the
%money options, or near dated options.
\end{comment}
%%%%%%%%%%%%%%%%%%%%%%%%%%%%%%%%%%%%%%%%%%%%%%%%%%%%%%%%%%%%%%%%%%%%%%%%%%%%%%%%%%%%%%%%%%%%%%%%%%%%%%%%%%%%%%%%%%
Additionally, in section \ref{Regime Switching and Pertubation
Based Option Pricing} it was proved that $\bar{\sigma}$ becomes an
increasingly better approximation of the stochastic volatility's
constant volatility equivalent as time increases. Consequently,
over shorter expiries $\bar{\sigma}$ is not as good an
approximation as the volatility process has not had sufficient
time to frequently
mean revert.% *p105-6: \cite{fouque2000dfm}p104-5

%Yang et al. \cite{yang:pco} compare option pricing methods over
%one quote date for a range of strikes and three expiries. We
%compare our three option pricing methods over eight different
%quote dates (four for each state), each over a range of expiries
%and strikes.

%Option data is not as easily available as equities data and has
%nor been in existence as long. Consequently, we were only able to
%test option pricing over the out of sample period of the
%calibrated RSexpOU model in chapter \ref{Chapter RSexpOU
%Stochastic Volatility Modelling and Calibration} (1997-2007).

\subsection{State 1 Option Pricing Results} We present
the results of our option pricing numerical experiments in state
1.
%NOT DATES BUT ERROR DIFFERENCE KEY\\
%FIND PAPER COMPARING OPTION PRICING METHODS -justify comparing few
%dates only.

\begin{table}[h!]
\begin{center}
\caption{Option Pricing Results for S\&P 500 Call
Option}\label{First Table SP500 State1}
\begin{tabular}{|c|c|c|c|c|c|}
\hline Expiry & Strike & Empirical  & \multicolumn{3}{|c|}{Option
Pricing Method}\\\cline{4-6}  &  & Price  & Black-Scholes &
\multicolumn{2}{|c|}{Perturbation Method} \tabularnewline
\cline{5-6}
 (Years) & (Cents) & (Cents) & Pricing & Standard & Regime Based \tabularnewline \hline
0.14 & 890 & 43.3 & 35.6 & 45.5 & 47.0\tabularnewline 0.14 & 895 &
39.9 & 31.9 & 40.7 & 41.7\tabularnewline 0.14 & 900 & 36.6 & 28.5
& 36.4 & 39.5\tabularnewline 0.14 & 910 & 30.4 & 22.3 & 28.8 &
28.9\tabularnewline 0.14 & 925 & 22.4 & 14.7 & 20.0 &
20.0\tabularnewline 0.14 & 940 & 15.7 & 9.1 & 13.7 &
14.2\tabularnewline 0.14 & 945 & 13.8 & 7.6 & 12.2 &
12.9\tabularnewline 0.21 & 900 & 43.2 & 33.0 & 42.0 &
42.5\tabularnewline 0.21 & 925 & 29.1 & 19.4 & 26.3 &
26.3\tabularnewline 0.38 & 900 & 54.4 & 41.1 & 52.1 &
52.6\tabularnewline 0.38 & 925 & 40.8 & 27.7 & 37.0 &
37.0\tabularnewline 0.63 & 900 & 66.3 & 50.5 & 64.2 &
64.7\tabularnewline 0.63 & 925 & 52.9 & 37.3 & 49.5 &
49.5\tabularnewline  \hline  &  & Average Percentage  & & &
\tabularnewline
 & & Error & 28.8\% & 6.9\% & 6.4\% \tabularnewline \hline
\end{tabular}
\flushleft{
\begin{small}
Quote date: 29/4/03
\end{small}
}
\end{center}
\end{table}

\newpage

\begin{table}[h!]
\begin{center}
\caption{Option Pricing Results for S\&P 500 Call
Option}\label{???}
\begin{tabular}{|c|c|c|c|c|c|}
\hline Expiry & Strike & Empirical  & \multicolumn{3}{|c|}{Option
Pricing Method}\\\cline{4-6}  &  & Price  & Black-Scholes &
\multicolumn{2}{|c|}{Perturbation Method} \tabularnewline
\cline{5-6}
 (Years) & (Cents) & (Cents) & Pricing & Standard & Regime Based \tabularnewline \hline
0.11 & 975 & 41.1 & 36.0 & 40.3 & 41.7\tabularnewline 0.11 & 980 &
37.9 & 32.2 & 35.9 & 36.7\tabularnewline 0.11 & 995 & 28.2 & 22.3
& 24.4 & 24.6\tabularnewline 0.11 & 1005 & 22.9 & 16.9 & 18.3 &
18.3\tabularnewline 0.11 & 1025 & 13.4 & 8.8 & 9.0 &
9.3\tabularnewline 0.20 & 975 & 49.9 & 41.6 & 45.6 &
46.5\tabularnewline 0.20 & 980 & 46.6 & 38.2 & 41.7 &
42.4\tabularnewline 0.20 & 985 & 43.4 & 35.0 & 38.0 &
38.5\tabularnewline 0.20 & 995 & 37.4 & 28.9 & 31.4 &
31.5\tabularnewline 0.20 & 1005 & 31.9 & 23.6 & 25.5 &
25.5\tabularnewline 0.20 & 1025 & 22.0 & 15.0 & 16.1 &
16.3\tabularnewline 0.45 & 1030 & 19.9 & 13.3 & 14.1 &
14.5\tabularnewline 0.45 & 975 & 64.9 & 53.4 & 57.6 &
58.3\tabularnewline 0.45 & 995 & 52.9 & 41.5 & 44.8 &
44.9\tabularnewline 0.45 & 1005 & 47.9 & 36.3 & 39.2 &
39.2\tabularnewline 0.45 & 1025 & 38.0 & 27.2 & 29.5 &
29.6\tabularnewline \hline
 &  & Average Percentage  &  &  & \tabularnewline
 &  & Error  & 23.0\%  & 16.5\%  & 15.6\% \tabularnewline
\hline
\end{tabular}
\flushleft{
\begin{small}
Quote date: 7/7/03
\end{small}
}
\end{center}
\end{table}

\newpage

\begin{table}[h!]
\begin{center}
\caption{Option Pricing Results for S\&P 500 Call
Option}\label{???}
\begin{tabular}{|c|c|c|c|c|c|}
\hline Expiry & Strike & Empirical  & \multicolumn{3}{|c|}{Option
Pricing Method}\\\cline{4-6}  &  & Price  & Black-Scholes &
\multicolumn{2}{|c|}{Perturbation Method} \tabularnewline
\cline{5-6}
 (Years) & (Cents) & (Cents) & Pricing & Standard & Regime Based \tabularnewline \hline 0.08 & 995 & 41.5 & 32.5 &
43.4 & 45.1\tabularnewline 0.08 & 1005 & 34.6 & 25.2 & 33.1 &
33.7\tabularnewline 0.08 & 1010 & 31.3 & 22.0 & 28.4 &
29.0\tabularnewline 0.08 & 1015 & 28.3 & 19.0 & 24.8 &
24.9\tabularnewline 0.08 & 1025 & 22.7 & 13.7 & 18.3 &
18.3\tabularnewline 0.08 & 1035 & 17.8 & 9.6 & 13.1 &
13.3\tabularnewline 0.08 & 1050 & 11.6 & 5.1 & 7.3 &
8.3\tabularnewline 0.17 & 995 & 50.4 & 38.8 & 49.8 &
50.9\tabularnewline 0.17 & 1005 & 44.0 & 32.2 & 41.4 &
41.9\tabularnewline 0.17 & 1025 & 32.4 & 21.2 & 28.2 &
28.2\tabularnewline 0.17 & 1050 & 20.4 & 11.4 & 17.0 &
17.6\tabularnewline 0.25 & 995 & 56.4 & 43.2 & 55.0 &
55.9\tabularnewline 0.25 & 1005 & 49.8 & 36.9 & 47.2 &
47.6\tabularnewline 0.25 & 1025 & 38.3 & 26.0 & 34.5 &
34.5\tabularnewline 0.25 & 1050 & 25.8 & 15.7 & 23.1 &
23.5\tabularnewline \hline
 &  & Average Percentage  &  &  & \tabularnewline
 &  & Error  & 33.5\%  & 11.8\%  & 10.7\% \tabularnewline
\hline
\end{tabular}
\flushleft{
\begin{small}
Quote date: 2/9/03
\end{small}
}
\end{center}
\end{table}

\newpage
\begin{table}[h!]
\begin{center}
\caption{Option Pricing Results for S\&P 500 Call
Option}\label{???}
\begin{tabular}{|c|c|c|c|c|c|}
\hline Expiry & Strike & Empirical  & \multicolumn{3}{|c|}{Option
Pricing Method}\\\cline{4-6}  &  & Price  & Black-Scholes &
\multicolumn{2}{|c|}{Perturbation Method} \tabularnewline
\cline{5-6}
 (Years) & (Cents) & (Cents) & Pricing & Standard & Regime Based \tabularnewline \hline
  0.12 & 1005 & 46.4 & 34.2 &
52.5 & 53.9\tabularnewline 0.12 & 1010 & 42.9 & 30.8 & 46.8 &
47.6\tabularnewline 0.12 & 1015 & 39.5 & 27.5 & 41.7 &
42.2\tabularnewline 0.12 & 1020 & 36.5 & 24.5 & 37.1 &
37.3\tabularnewline 0.12 & 1025 & 33.2 & 21.7 & 32.9 &
33.0\tabularnewline 0.12 & 1050 & 19.5 & 10.7 & 17.0 &
17.4\tabularnewline 0.12 & 1060 & 15.6 & 7.8 & 12.2 &
13.1\tabularnewline 0.20 & 1005 & 53.1 & 39.2 & 58.3 &
59.3\tabularnewline 0.20 & 1015 & 46.7 & 32.8 & 48.9 &
49.3\tabularnewline 0.20 & 1025 & 40.5 & 27.1 & 40.9 &
41.0\tabularnewline 0.20 & 1035 & 34.7 & 22.1 & 34.1 &
34.1\tabularnewline 0.20 & 1050 & 27.4 & 15.9 & 25.7 &
26.1\tabularnewline 0.20 & 1055 & 25.0 & 14.1 & 23.3 &
23.8\tabularnewline 0.20 & 1060 & 22.6 & 12.5 & 21.1 &
21.8\tabularnewline \hline
 &  & Average Percentage  &  &  & \tabularnewline
 &  & Error  & 36.0\%  & 7.2\%  & 6.9\% \tabularnewline
\hline
\end{tabular}
\flushleft{
\begin{small}
Quote date: 6/10/03
\end{small}
}
\end{center}
\end{table}

\newpage
\begin{figure}[h!]
\begin{center}
\caption{Graph of Table \ref{First Table SP500 State1} for Options
with Expiry of 0.14 Years}\label{Graph for Table for Options with
Expiry of 0.14 Years}
\includegraphics[height=13cm]{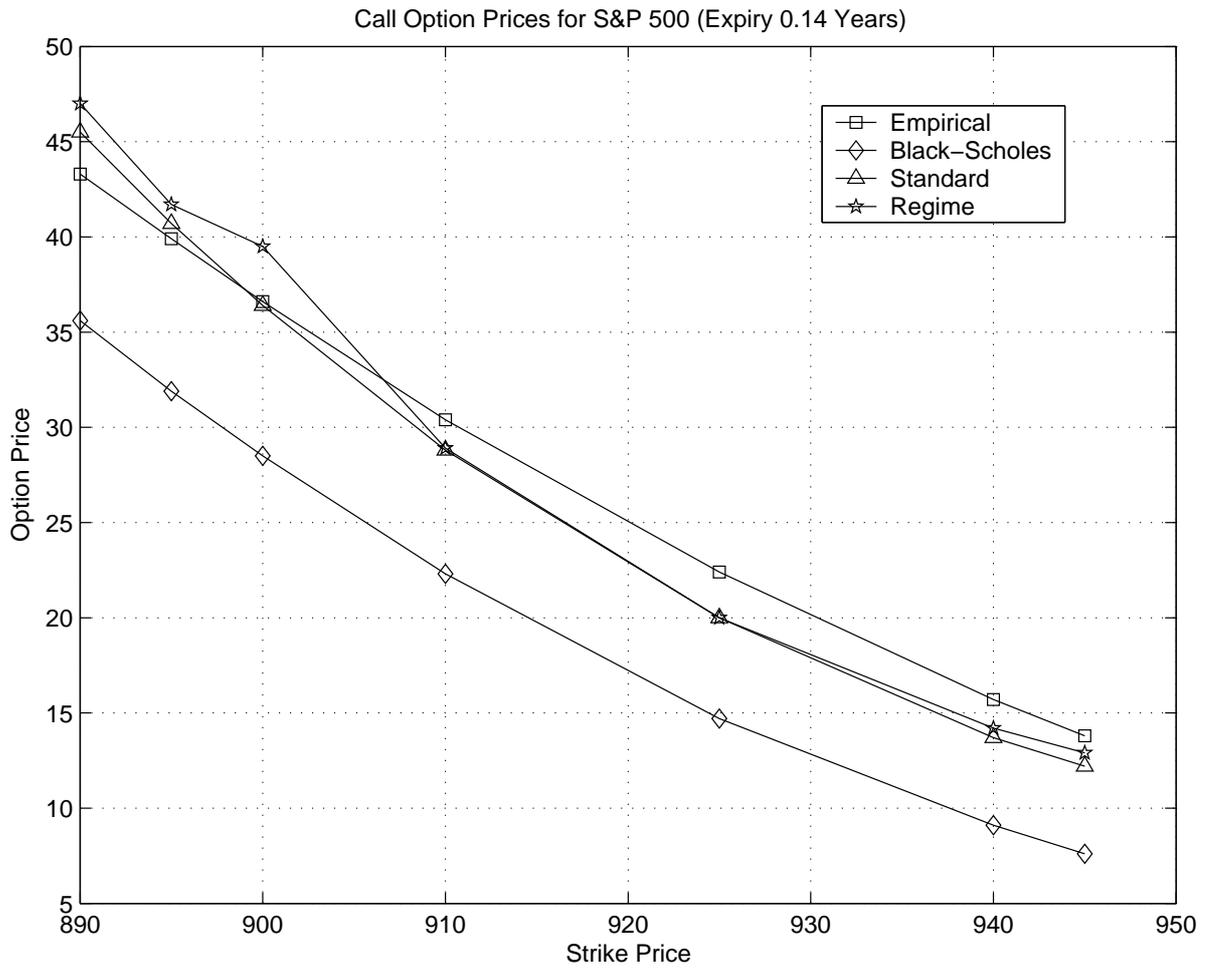}
\end{center}
\end{figure}

\newpage
%\subsection*{}%\\
The numerical experiments give option pricing results for state
one, from four different quote periods over a range of strikes and
expiries. We see that regime based option pricing provides lower
average percentage error compared to Black-Scholes or standard
Fouque option pricing. In fact the average Black-Scholes option
pricing error is 30.3\% whereas Fouque's standard and regime based
average option pricing errors are 10.6\% and 9.9\% respectively.

The average Black-Scholes option pricing error of 30.3\% over
different expiries and strikes validates Fouque's perturbation
solution to stochastic volatility option pricing. Since the option
price C is solved by the perturbation expansion equation (\ref{C
pert exp eqn})
%\begin{eqnarray}
%C=C_{0}+\sqrt{\epsilon}C_{1}+\epsilon
%C_{2}+{\epsilon}^{\frac{3}{2}}C_{3}+.... ,
%\end{eqnarray}
 therefore C should be approximately equal to the leading order
term $C_{0}$, which is the Black-Scholes option price.
Additionally, it supports our claim that $C_{0}$ can be used as an
approximation for intra-regime option prices.

Figure \ref{Graph for Table for Options with Expiry of 0.14 Years}
plots table \ref{First Table SP500 State1} for options with expiry
of 0.14 years, with the empirical option prices labelled
``empirical".
%DON'T MENTION BELOW -SV SMILE WITH STRIKE SO INCORRECT
% are consistent with theoretical expectations, for
%instance the option prices decrease with strike. This suggests
%this empirical set of option prices are priced by ``rational"
%rather than ``irrational" factors and so form a feasible data set
%for analysing results.
It is reassuring that all three pricing methods follow the same
trend as empirical option prices, providing evidence that all
three are viable option pricing methods. We observe that the
Black-Scholes option prices consistently underprice over all
strikes compared to the empirical prices, which has been
frequently observed by many researchers. For instance, the
Black-Scholes underpricing is most easily noticeable when
empirical option prices exhibit volatility smiles.

The Black-Scholes underpricing has been explained by researchers
recognising that Black-Scholes option pricing assumes constant
volatility and therefore no risk premium can be associated with
any ``volatility risk". In stochastic volatility our risk premium
associated with volatility is accounted for by the market price of
volatility risk term, therefore it is possible to price options
with volatility risk. In Fouque's perturbation solution to option
pricing, $C_{1}$ contains the market price of volatility risk,
therefore suffers less from underpricing compared to the
Black-Scholes equation.

%%%%%%%%%%%%%%%%%%%%%%%%%%%%%%%%%%%%%%%%%%%%%%%%%%%%%%%%%%%%%%%%%%%%%%%%%%%%%%%%%%%%%%%%%%%%%%%%%%%%%%%%%%%%%%
\begin{comment}

Although the regime based option prices are not substantially
different from standard Fouque option pricing this is not
unexpected. From chapter \ref{Chapter RSexpOU Stochastic
Volatility Modelling and Calibration} tables \ref{In Sample Regime
Sequence Results} and \ref{Out of Sample Regime Sequence Results}
we can see the RSexpOU process spends the majority of the time in
state one. Consequently the state one RSexpOU dynamics should be
similar to the non-regime switching expOU dynamics; this is
confirmed by the results obtained in tables \ref{RSexpOU
Parameters} and \ref{expOU Parameters (no regime switching)}.
Therefore we would not expect a substantial improvement in option
pricing in state one.

propose offers beeter option pricing since regime specific vol
takes into account current economic environment so vol more
relevant

modelling underlier better not ncessarily improve option pricing
since some contend options asset different to underlier altogether
schonbucher model.

\end{comment}
%%%%%%%%%%%%%%%%%%%%%%%%%%%%%%%%%%%%%%%%%%%%%%%%%%%%%%%%%%%%%%%%%%%%%%%%%%%%%%%%%%%%%%%%%%%%%%%%%%%%%%%%%%%%%%

\newpage
\subsection{State 2 Option Pricing Results}
We present the results of our option pricing numerical experiments
in state two.
\begin{table}[htbp!]
\begin{center}
\caption{Option Pricing Results for S\&P 500 Call
Option}\label{First Table SP500 State 2}
\begin{tabular}{|c|c|c|c|c|c|}
\hline Expiry & Strike & Empirical  & \multicolumn{3}{|c|}{Option
Pricing Method}\\\cline{4-6}  &  & Price  & Black-Scholes &
\multicolumn{2}{|c|}{Perturbation Method} \tabularnewline
\cline{5-6}
 (Years) & (Cents) & (Cents) & Pricing & Standard & Regime Based \tabularnewline \hline
0.12 & 1250 & 75.6 & 49.7 & 94.3 & 92.8\tabularnewline 0.12 & 1275
& 59.4 & 32.5 & 58.4 & 58.2\tabularnewline 0.12 & 1300 & 44.6 &
19.4 & 35.7 & 35.6\tabularnewline 0.21 & 1250 & 89.9 & 59.5 &
106.5 & 105.2\tabularnewline 0.21 & 1275 & 73.9 & 42.7 & 75.1 &
74.8\tabularnewline 0.21 & 1285 & 68.0 & 36.9 & 65.4 &
65.3\tabularnewline 0.21 & 1300 & 59.1 & 29.2 & 53.1 &
53.1\tabularnewline 0.45 & 1250 & 122.6 & 84.1 & 143.7 &
142.0\tabularnewline 0.45 & 1275 & 106.9 & 67.6 & 115.3 &
114.7\tabularnewline 0.45 & 1300 & 91.8 & 53.2 & 93.1 &
93.0\tabularnewline 0.72 & 1250 & 150.5 & 107.0 & 180.2 &
178.0\tabularnewline 0.72 & 1275 & 135.0 & 90.5 & 152.1 &
151.0\tabularnewline 0.72 & 1300 & 120.0 & 75.6 & 129.1 &
128.7\tabularnewline 0.97 & 1300 & 145.4 & 95.1 & 160.0 &
159.2\tabularnewline \hline
 &  & Average Percentage  & 39.4\% & 11.2\% & 10.6\%\tabularnewline
 &  & Error  &  &  &
\tabularnewline \hline
\end{tabular}
\end{center}
\flushleft{
\begin{small}
Quote date: 2/1/01
\end{small}
}
\end{table}

\newpage

\begin{table}[h!]
\begin{center}
\caption{Option Pricing Results for S\&P 500 Call Option
}\label{??}
\begin{tabular}{|c|c|c|c|c|c|}
\hline Expiry & Strike & Empirical  & \multicolumn{3}{|c|}{Option
Pricing Method}\\\cline{4-6}  &  & Price  & Black-Scholes &
\multicolumn{2}{|c|}{Perturbation Method} \tabularnewline
\cline{5-6}
 (Years) & (Cents) & (Cents) & Pricing & Standard & Regime Based \tabularnewline \hline

0.14 & 1200 & 66.9 & 50.7 & 82.5 & 81.3\tabularnewline 0.14 & 1210
& 60.1 & 43.5 & 68.5 & 67.8\tabularnewline 0.14 & 1225 & 50.6 &
33.8 & 51.7 & 51.5\tabularnewline 0.14 & 1250 & 36.9 & 20.8 & 31.3
& 31.3\tabularnewline 0.23 & 1200 & 80.2 & 60.7 & 93.4 &
92.3\tabularnewline 0.23 & 1225 & 64.3 & 44.3 & 66.3 &
66.1\tabularnewline 0.23 & 1250 & 50.3 & 30.9 & 46.7 &
46.7\tabularnewline 0.40 & 1200 & 98.5 & 75.6 & 112.8 &
111.7\tabularnewline 0.40 & 1225 & 83.0 & 59.6 & 87.9 &
87.5\tabularnewline 0.40 & 1250 & 68.9 & 45.8 & 68.5 &
68.4\tabularnewline 0.65 & 1225 & 105.4 & 78.9 & 112.8 &
114.5\tabularnewline \hline
 &  & Average Percentage  & 32.1\% & 9.6\% & 9.0\%\tabularnewline
 &  & Error  &  &  & \tabularnewline
\hline
\end{tabular}
\end{center}
\flushleft{
\begin{small}
Quote date: 26/4/01
\end{small}
}
\end{table}
\newpage

\begin{table}[h!]
\begin{center}
\caption{Option Pricing Results for S\&P 500 Call Option}
\begin{tabular}{|c|c|c|c|c|c|}
\hline Expiry & Strike & Empirical  & \multicolumn{3}{|c|}{Option
Pricing Method}\\\cline{4-6}  &  & Price  & Black-Scholes &
\multicolumn{2}{|c|}{Perturbation Method} \tabularnewline
\cline{5-6}
 (Years) & (Cents) & (Cents) & Pricing & Standard & Regime Based \tabularnewline \hline
0.08 & 1175 & 50.9 & 41.1 & 62.3 & 61.2\tabularnewline 0.08 & 1200
& 32.9 & 23.6 & 34.2 & 34.1\tabularnewline 0.08 & 1225 & 20.5 &
11.6 & 17.5 & 17.5\tabularnewline 0.17 & 1175 & 64.4 & 50.5 & 71.6
& 70.8\tabularnewline 0.17 & 1190 & 55.0 & 40.2 & 56.7 &
56.4\tabularnewline 0.17 & 1200 & 47.1 & 34.1 & 48.5 &
46.3\tabularnewline 0.17 & 1210 & 41.3 & 28.6 & 41.3 &
41.3\tabularnewline 0.17 & 1225 & 33.8 & 21.5 & 32.4 &
32.3\tabularnewline 0.17 & 1240 & 26.9 & 15.7 & 25.3 &
25.1\tabularnewline 0.42 & 1175 & 90.3 & 70.1 & 96.7 &
96.0\tabularnewline  \hline
 &  & Average Percentage  & 30.0\% & 7.3\% & 6.9\%\tabularnewline
 &  & Error  &  &  & \tabularnewline\hline
\end{tabular}
\end{center}
\flushleft{
\begin{small}
Quote date: 18/7/01
\end{small}
}
\end{table}

\newpage

\begin{table}[h!]
\begin{center}
\caption{Option Pricing Results for S\&P 500 Call
Option}\label{Last Table SP500 State 2}
\begin{tabular}{|c|c|c|c|c|c|}
\hline Expiry & Strike & Empirical  & \multicolumn{3}{|c|}{Option
Pricing Method}\\\cline{4-6}  &  & Price  & Black-Scholes &
\multicolumn{2}{|c|}{Perturbation Method} \tabularnewline
\cline{5-6}
 (Years) & (Cents) & (Cents) & Pricing & Standard & Regime Based \tabularnewline \hline
0.11 & 1060 & 63.4 & 40.6 & 80.6 & 79.2\tabularnewline 0.11 & 1070
& 56.7 & 33.2 & 64.3 & 63.7\tabularnewline 0.11 & 1075 & 53.3 &
29.8 & 57.7 & 57.3\tabularnewline 0.11 & 1080 & 50.2 & 26.6 & 51.8
& 51.6\tabularnewline 0.11 & 1090 & 44.1 & 20.9 & 42.1 &
42.0\tabularnewline 0.11 & 1100 & 37.8 & 15.9 & 34.4 &
34.4\tabularnewline 0.11 & 1125 & 26.3 & 7.3 & 21.7 &
21.2\tabularnewline 0.18 & 1100 & 48.7 & 22.3 & 47.2 &
47.2\tabularnewline 0.18 & 1125 & 36.0 & 12.6 & 34.6 &
34.2\tabularnewline 0.28 & 1160 & 83.4 & 52.3 & 97.2 &
96.4\tabularnewline 0.28 & 1100 & 59.0 & 29.0 & 60.5 &
60.5\tabularnewline 0.28 & 1125 & 46.1 & 18.6 & 47.5 &
47.3\tabularnewline \hline
 &  & Average Percentage  &  &  & \tabularnewline
 &  & Error  & 51.5\%  & 9.4\%  & 9.1\% \tabularnewline
\hline
\end{tabular}
\end{center}
\flushleft{
\begin{small}
Quote date: 10/9/01
\end{small}
}
\end{table}

\newpage
\begin{figure}[h!]
\begin{center}
\caption{Graph of Table \ref{Last Table SP500 State 2} for Options
with Expiry of 0.11 Years}\label{Graph for Table State 2 for
Options with Expiry of 0.11 Years}
\includegraphics[height=13cm]{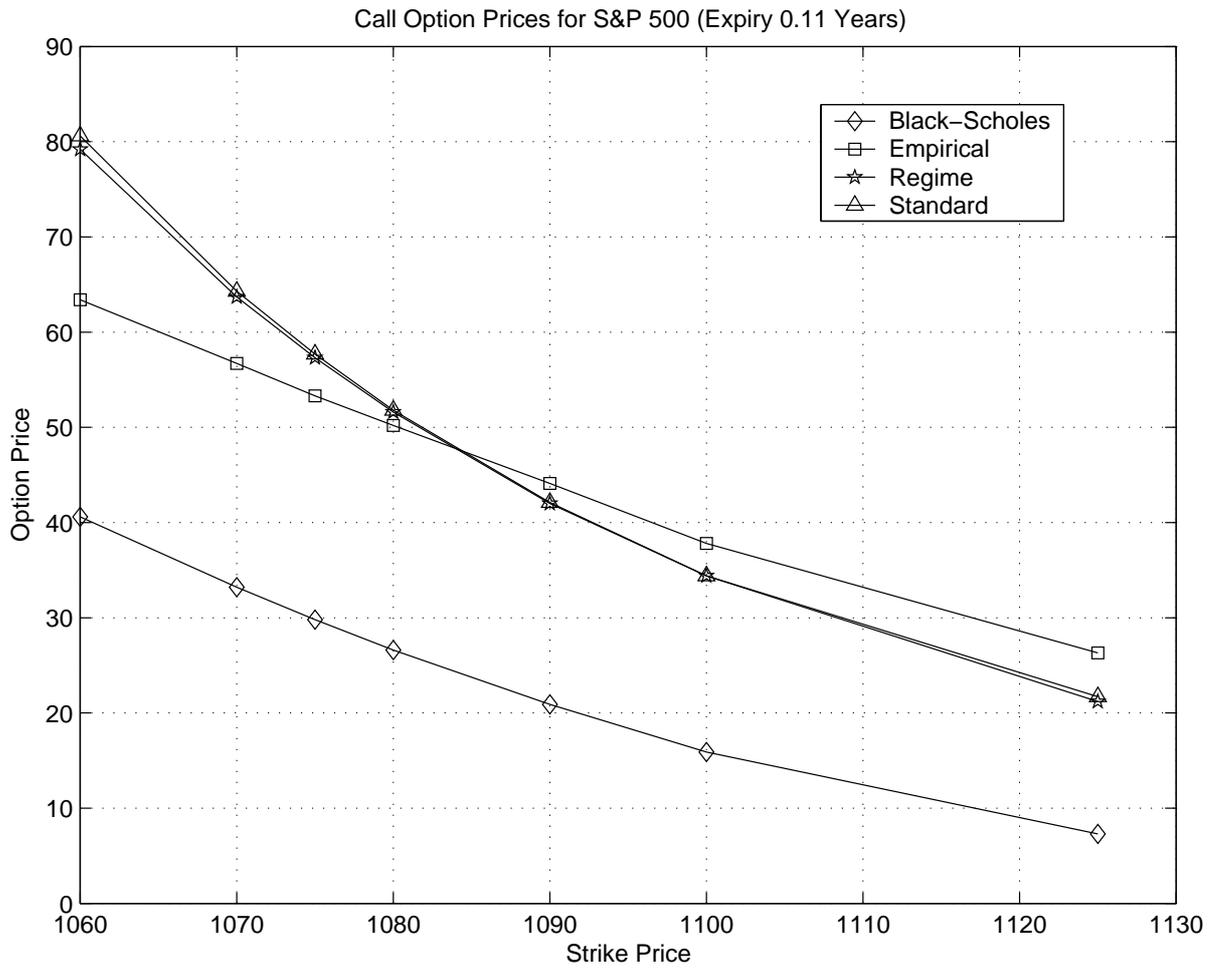} %[height=10cm]
\end{center}
\end{figure}
\newpage
%\textbf{Discussion}\\
The numerical experiments give option pricing results for state
two, from four different quote periods over a range of strikes and
expiries. As in state one regime based option pricing provides
lower average percentage error compared to Black-Scholes or
standard Fouque option pricing. In fact the average Black-Scholes
option pricing error is 38.2\% whereas Fouque's standard and
regime based average option pricing errors are 9.4\% and 8.9\%
respectively. Therefore the pricing error under Black-Scholes
option pricing is higher in state two than in state one, which is
consistent with empirical observations as volatility smiles
increase during ``down states" (state two) and therefore option
prices deviate away from a constant volatility assumption.
%%%%%%%%%%%%%%%%%%%%%%%%%%%%%%%%%%%%%%%%%%%%%%%%%%%%%%%%%%%%%%%%%%%%%%%%%%%%%%%%%%%%%%%%%%%%%%%%%%%%%%%%%%%%%%%%%
\begin{comment}
From the tables' results the difference in average percentage
error between regime based and non-regime based option pricing
over all strikes and expiries is greater in state 2 than compared
to state one. From chapter \ref{Chapter RSexpOU Stochastic
Volatility Modelling and Calibration} tables \ref{In Sample Regime
Sequence Results} and \ref{Out of Sample Regime Sequence Results}
we can see RSexpOU process spends the minority of the time in
state 2. Consequently the state 2 RSexpOU dynamics are not similar
to the non-regime switching expOU dynamics; this is confirmed by
the results obtained in tables \ref{RSexpOU Parameters} and
\ref{expOU Parameters (no regime switching)}. Therefore we would
expect a greater difference in option prices, which is confirmed
by the difference in pricing error in this section.
\end{comment}
%%%%%%%%%%%%%%%%%%%%%%%%%%%%%%%%%%%%%%%%%%%%%%%%%%%%%%%%%%%%%%%%%%%%%%%%%%%%%%%%%%%%%%%%%%%%%%%%%%%%%%%%%%%%%%%%%

Figure  \ref{Graph for Table State 2 for Options with Expiry of
0.11 Years} plots table \ref{First Table SP500 State 2}  for
options with expiry of 0.11 years with the empirical option prices
labelled ``empirical".
%DON'T MENTION -THEORETICALLY UNDER SV PRICES SMILE
%As in state 1, we observe that the empirical option prices
%(labelled ``empirical") are consistent with theoretical
%expectations, suggesting the prices are governed by ``rational"
%rather than ``irrational" factors. Therefore the empirical data
%form a feasible set for analysing results.
As in state one, all three pricing methods follow the same trend
as empirical option prices, providing evidence that all three are
viable option pricing methods in state two. We also note the
Black-scholes option prices are consistently underpricing over all
strikes, as in state one, for the same reasons as in state one
-Black-Scholes option pricing does not take into account
``volatility risk".

From figure \ref{Graph for Table State 2 for Options with Expiry
of 0.11 Years} we observe that both Fouque based option pricing
methods exhibit a more pronounced volatility smiling effect
compared to figure \ref{Graph for Table for Options with Expiry of
0.14 Years}. Since we price options upto the first correction
$\sqrt{\epsilon}C_{1}$ and $C_{0}$ does not take into account any
volatility smiling effects, we can attribute the smiling effect to
$\sqrt{\epsilon}C_{1}$. The term $\sqrt{\epsilon}C_{1}$ can
increase volatility smiling through an increase in the market
price of volatility risk, which is likely because state two models
the ``down" state and we expect risk associated with volatility to
be higher. Additionally, we notice that $\sqrt{\epsilon}C_{1}$
contains $\rho$, the correlation between the volatility's Wiener
process $W_{2}$ and $W_{1}$, which may increase in ``down" states
compared to ``up" states. This is consistent with empirical
observations since volatility tends to increase during a down
state.

\section{Conclusions}
It can be seen from our literature review of volatility models
that the development of volatility models has progressed in a
logical order to address key shortcomings of previous models. Time
dependent models addressed option prices varying with expiration
dates, local volatility also addressed volatility smiles and the
leverage effect, whereas stochastic volatility could incorporate
all the effects captured by local volatility and a range of other
empirical effects e.g. greater variability in observed volatility.
However the trade-off associated with improved volatility
modelling has been loss of analytical tractability.

In conclusion we have shown that Fouque's option pricing method
provides significantly better option pricing accuracy compared to
Black-Scholes option pricing over a variety of strikes and
expiries. We have provided option prices and have shown that
taking into account regime specific $\bar{\sigma}_{i}$, rather
than using one $\bar{\sigma}$, improves option pricing accuracy.

\newpage
\bibliographystyle{plain}
\addcontentsline{toc}{section}{References}
\bibliography{Ref}

\end{document}